\documentclass[english,aps,prb,amsmath,tightenlines,notitlepage]{revtex4-1}
\usepackage[T1]{fontenc}
\usepackage[latin9]{inputenc}
\usepackage{geometry}
\usepackage{color}
\usepackage{amsmath}
\usepackage{amsthm}
\usepackage{amssymb}
\usepackage{graphicx}
\usepackage{hyperref}
\usepackage{comment}

\makeatletter

\providecommand{\tabularnewline}{\\}

\numberwithin{equation}{section}
\numberwithin{figure}{section}

\usepackage{babel}
\begin{document}
\global\long\def\ve{\varepsilon}%
\global\long\def\R{\mathbb{R}}%
\global\long\def\Rn{\mathbb{R}^{n}}%
\global\long\def\Rd{\mathbb{R}^{d}}%
\global\long\def\E{\mathbb{E}}%
\global\long\def\P{\mathbb{P}}%
\global\long\def\bx{\mathbf{x}}%
\global\long\def\vp{\varphi}%
\global\long\def\ra{\rightarrow}%
\global\long\def\smooth{C^{\infty}}%
\global\long\def\Tr{\mathrm{Tr}}%
\global\long\def\bra#1{\left\langle #1\right|}%
\global\long\def\ket#1{\left|#1\right\rangle }%
\global\long\def\Re{\mathrm{Re}}%
\global\long\def\Im{\mathrm{Im}}%
\global\long\def\bsig{\boldsymbol{\sigma}}%
\global\long\def\btau{\boldsymbol{\tau}}%
\global\long\def\bmu{\boldsymbol{\mu}}%
\global\long\def\bx{\boldsymbol{x}}%
\global\long\def\bups{\boldsymbol{\upsilon}}%
\global\long\def\bSig{\boldsymbol{\Sigma}}%
\global\long\def\bt{\boldsymbol{t}}%
\global\long\def\bG{\boldsymbol{G}}%
\global\long\def\bB{\boldsymbol{B}}%
\global\long\def\bs{\boldsymbol{s}}%
\global\long\def\by{\boldsymbol{y}}%
\global\long\def\brho{\boldsymbol{\rho}}%
\global\long\def\ba{\boldsymbol{a}}%
\global\long\def\bb{\boldsymbol{b}}%
\global\long\def\bz{\boldsymbol{z}}%
\global\long\def\bc{\boldsymbol{c}}%
\global\long\def\balpha{\boldsymbol{\alpha}}%
\global\long\def\bbeta{\boldsymbol{\beta}}%
\global\long\def\br{\boldsymbol{r}}%
\global\long\def\bj{\boldsymbol{j}}%
\global\long\def\bi{\boldsymbol{i}}%
\global\long\def\bk{\boldsymbol{k}}%
\global\long\def\bL{\boldsymbol{L}}%
\global\long\def\bn{\boldsymbol{n}}%
\global\long\def\bm{\boldsymbol{m}}%
\global\long\def\T0{\mathbf{T}}%
\global\long\def\Tt{\tilde{\mathbf{T}}}%
\global\long\def\Lap{\mathbf{L}}%
\global\long\def\Lt{\tilde{\mathbf{L}}}%

\title{Fast and spectrally accurate construction of \\ adaptive diagonal basis sets for electronic structure}
\author{Michael Lindsey}
\email{lindsey@math.berkeley.edu }

\affiliation{Department of Mathematics, University of California, Berkeley, Berkeley,
CA 94720, USA}
\author{Sandeep Sharma}
\email{sanshar@gmail.com}

\affiliation{Department of Chemistry, University of Colorado, Boulder, CO 80302,
USA}
\begin{abstract}
In this article, we combine the periodic sinc basis set with a curvilinear coordinate system for electronic structure calculations. This extension allows for variable resolution across the computational domain, with higher resolution close to the nuclei and lower resolution in the inter-atomic regions. We address two key challenges that arise while using basis sets obtained by such a coordinate transformation. First, we use pseudospectral methods to evaluate the integrals needed to construct the Hamiltonian in this basis. Second, we demonstrate how to construct an appropriate coordinate transformation by solving the Monge-Amp\`ere equation using a new approach that we call the cyclic Knothe-Rosenblatt flow. The solution of both of these challenges enables mean-field calculations at a cost that is log-linear in the number of basis functions. We demonstrate that our method approaches the complete basis set limit faster than basis sets with uniform resolution. We also emphasize how these basis sets satisfy the diagonal approximation, which is shown to be a consequence of the pseudospectral method. The diagonal approximation is highly desirable for the solution of the electronic structure problem in many frameworks, including mean field theories, tensor network methods, quantum computing, and quantum Monte Carlo.

\end{abstract}
\maketitle

\section{Introduction}

The first step in the solution of the electronic structure problem
involves the discretization of the Hamiltonian using a basis set.
Two types of basis sets are most commonly used: first, atom-centered 
basis functions (with Gaussian type orbitals (GTO) being
the most common in this category\citep{molpro,molcas,g16,orca,dalton,dirac,nwchem,gamess,psi4,pyscf,cp2k,turbomole,crystal},
although numerical atomic orbitals are also used\citep{fhi-aims})
and, second, plane wave basis sets\citep{quantum_espresso,vasp}.
These basis sets have complementary strengths. With GTOs one is able
to perform all-electron calculations because they are well-suited
to describing the nuclear cusp in the wavefunction. They usually deliver
impressive accuracy even when very few basis functions are used, i.e., roughly $O(10)$
per atom, although converging results to the complete basis set limit
with error smaller than $\mu$E$_{h}$ can be challenging. Plane wave
(PW) basis functions, on the other hand, usually require the use of a
pseudopotential (PP) as well as a much larger number of basis
functions, roughly $O(1000)$ per atom when PPs are used. However,
plane waves enjoy spectral convergence to the exact result (within
the PP approximation), and the complete basis set limit can be
achieved routinely. In addition, the various integrals in the Hamiltonian
take on a particularly simple form with PW. Moreover, the use
of fast Fourier transforms (FFTs) enables mean-field
calculations with a computational cost that scales better than mean-field calculations 
with GTOs by a factor proportional to the system size, without further approximations such
as tensor hypercontraction (THC)\cite{THC1,THC2}. The disadvantage of the PW basis is that
it is not adapted to the molecular geometry. As a result, PW calculations require uniformly
high resolution across the entire computational domain, unlike GTOs which
provide high resolution close to nuclear centers and lower resolution
in the interatomic space. 

More recently, other constructions including multiwavelet\citep{madness,bigdft,mrchem}, finite difference\cite{octopus,gpaw}, and 
finite element methods\citep{GaviniEnriched1,rufus2021fast,dft-fe-1.0,lin_dg} have risen to prominence, 
offering variable resolution over different
parts of the computational domain and enabling
all-electron calculations with guaranteed accuracy. Some of the most
accurate, benchmark-quality all-electron calculations for large system
sizes have been performed using these basis sets. However, to the best of the authors' knowledge, such approaches have not yet been used to write the full electronic Hamiltonian in second quantization. This makes it difficult to use
them using general correlated calculations, although we note that perturbation
theory calculations using multiwavelet basis sets are feasible\citep{kottmann2020}. These approaches also do not deliver spectral convergence with increasing basis set size, in contrast with the PW basis.

In 1992, Gygi introduced a PW basis defined with respect to a curvilinear coordinate
system\citep{Gygi1992,Gygi1993,Gygi1995} that deforms
the domain such that more basis functions are concentrated around the
nucleus and fewer basis functions are included in the interatomic space. 
Such a perspective aims to build the adaptivity of an atom-centered basis 
into the PW framework, maintaining its benefits such as spectral accuracy and fast 
implementation. The curvilinear
coordinates were optimized to minimize the energy of the mean-field
calculations using a double loop: in the inner loop the coefficients
of the molecular orbitals were optimized, and in the outer loop the
map describing the curvilinear coordinates was optimized. This work
inspired other followup calculations by various groups\citep{Devenyi1994,JoseM.Perez-Jorda1995,Perez-Jorda1998,Perez-Jorda2011,Perez-Jorda2014,Hamann1995,Hamann1996,Hamann2001,Hamman1995,Modine1997,Zumbach1996,Rodriguez2008}.

Our work here is inspired by the pioneering work of Gygi, and we modify
the approach by incorporating the so-called diagonal approximation, as well 
as a new framework for computing the deformation.
The concept of the diagonal approximation was formalized in the work of 
White et al in the development of the Gausslet basis functions\citep{White2017} and various extensions \citep{White2019,Qiu2021}.
This work demonstrates that for basis functions which are orthogonal and interpolating
(i.e., behaving like delta functions,
cf. Section \ref{sec:prelim} for more details),
the two-electron integrals can be effectively replaced by a 
two-index, rather than four-index, quantity. The original Gausslets (like PWs)
offer uniform resolution across the entire computational domain and thus suffer
from the same problems as PW in terms of the efficiency of the basis representation. 
We highlight that the Gausslets were designed in part for deployment within 
tensor network approaches to electronic structure, where they nonetheless 
enjoy advantages over PWs in maintaining locality of entanglement and allowing for 
efficient representation of the many-body Hamiltonian. It is worth pointing out that such `diagonal' basis sets also deliver significant advantage for quantum computing applications\cite{babbush2018low,mcclean2020discontinuous}. 

All the same, the original shortcoming 
of non-adaptivity was partially overcome in the work of White and Stoudenmire\citep{White2019} by introducing
curvilinear coordinates while maintaining the diagonal approximation. The deformation
used in this work is of separable or multi-sliced (cf. Section~\ref{sec:reviewKR} below) form, 
enabling essentially exact evaluation of all required integrals. However, for a fully 
3-D system, multi-sliced deformations are not flexible enough to distribute basis functions 
ideally throughout the computational domain. This shortcoming was addressed in a 
recent paper by White and Lindsey\citep{White2023}, which is not simply based on a deformation 
of an underlying grid of basis functions. This work introduces nested Gausslets basis sets, 
which are carefully constructed using a procedure based on the coarsening of 1-D diagonal basis sets constructed 
in terms of a separable deformation near atom centers. In particular, the `COMX theorem' demonstrates that this 
coarsening can be achieved with a simple and tractable diagonalization. Ultimately the nested Gausslet construction eliminates unnecessary functions in regions where high resolution is not needed, delivering controlled accuracy with a small number of basis functions and maintaining the diagonal approximation. In addition, the construction 
can be combined with GTOs while still maintaining diagonality. However, so far the construction has only been realized for atomic and diatomic molecules, although extensions to fully 3-D systems may
be possible with suitable extensions.
A side effect of using nested Gausslets is that one
needs to store a dense matrix of two-electron integrals, with memory cost
scaling quadratically in the number of nested Gausslets. Although
the basis set size significantly improves on that of the original Gausslets, the relatively large
number of basis functions compared to a typical GTO basis set could make the storage of two-index quantities
a potential bottleneck for large systems.

In this work we combine the periodic sinc basis set\cite{boyd2001chebyshev} with a curvilinear coordinate system. 
The sinc functions are identical in span to plane waves but allow us to achieve the diagonal approximation, 
which we explain as a consequence of pseudospectral approximation \cite{boyd2001chebyshev}. 
Our curvilinear deformation
is fully 3-D, allowing for the treatment of general molecular geometries in a straightforward and automatic fashion. However, due to the use
of general 3-D deformations, it is nontrivial to calculate the integrals required to construct the Hamiltonian. Following the work of Gygi\cite{Gygi1992}, we demonstrate that they can be computed efficiently 
within a pseudospectral approximation framework that introduces little additional error on top of the basis 
truncation error itself. In addition, we eliminate the non-analyticity
due to nuclear cusp by using all-electron pseudopotentials of Gygi\citep{Gygi2023}
that are highly accurate and can be systematically improved to arbitrary
precision, enabling our pseudospectral approach to enjoy spectral convergence as the discretization is refined.  Using our implementation,
the cost of mean-field calculations scales log-linearly with the number
of basis functions, just like calculations in a traditional PW basis. However, in our approach 
the number of basis functions is dramatically reduced.

A novel aspect of our work is the computation of the map or deformation defining the curvilinear 
coordinate system. In contrast to preceding work, we take as given a prescribed density of 
grid points that the deformation should achieve. Such a density can be supplied either from an 
approximate preliminary calculation or from a formula inspired
by multiresolution analysis. In particular, the functional form that we have used was
suggested in the nested Gausslet paper\citep{White2023}, cf. \eqref{eq:deform} below. We shall
demonstrate that this approach based on matching a prescribed density
yields rapidly converging energies as the discretization is refined. Moreover, the invertibility 
of the deformation is guaranteed by the construction, even when the deformation is quite sharp. 
This property is difficult to maintain in more heuristic approaches when extreme deformations are required. Although the goal of matching a prescribed density poses an apparently
difficult computational problem, namely, the problem of solving a
Monge-Amp\`ere equation~\cite{Villani_2003}, we show how this problem can
be solved efficiently with spectral accuracy using a novel approach. We call our transformation
solving this problem the \emph{cyclic Knothe-Rosenblatt flow}, which can be viewed as a 
collocation method for the Monge-Amp\`ere equation in both forward and inverse formulations. In particular, we demonstrate that computing the cyclic 
Knothe-Rosenblatt flow achieving the map from uniform to curvilinear coordinates can be achieved with Chebyshev or PW functions at a cost that scales linearly with the number of function, so that the step of computing the deformation does not impose a computational bottleneck in the broader context of our work.

The rest of the paper is organized as follows. In Section 2, we begin
with preliminaries and briefly describe the pseudospectral method and diagonal approximation.
In Section 3, we show how the use of the pseudospectral method leads to a
diagonal approximation for our deformed sinc basis set. In Section 4, we
show how to solve the Monge-Amp\`ere equation using the cyclic
Knothe-Rosenblatt flow. In Section 5, we show how the integrals needed
for mean-field calculations are evaluated, and we also describe our
log-linear-scaling algorithm for solving the Hartree-Fock equations. Then in Section 6 
we present numerical results demonstrating that rapid energy convergence is obtained
for the various systems we have studied. Finally, we close with
conclusions and a discussion of future work. 

\section{Preliminaries \label{sec:prelim}}

Consider an electronic stucture problem, discretized with the orthonormal
basis set $\{\eta_{i}(\bx)\}$. The quantum chemistry Hamiltonian
\begin{equation}
\label{eq:secondquant}
\sum_{i,j}\mathbf{H}_{ij}\,a_{i}^{\dagger}a_{j}+\frac{1}{2}\sum_{ijkl}\mathbf{V}_{ijkl}\,a_{i}^{\dagger}a_{k}^{\dagger}a_{l}a_{j}
\end{equation}
 is specified by one-electron and two-electron integrals $\{\mathbf{H}_{ij}\}$
and $\{\mathbf{V}_{ijkl}\}$, respectively. In turn $\mathbf{H}_{ij}=\mathbf{T}_{ij}+\mathbf{U}_{ij}$
can be written as a sum of kinetic and potential contributions 
\begin{equation}
\mathbf{T}_{ij}=-\int\eta_{i}(\bx)\Delta\eta_{j}(\bx)\,d\bx,\quad\mathbf{U}_{ij}=\sum_{I}\int\eta_{i}(\bx)V_{I}(\bx)\eta_{j}(\bx)\,d\bx,\label{eq:1electron}
\end{equation}
 where $\Delta$ denotes the Laplacian operator, $I$ indexes the
atoms with atomic numbers $Z_{I}$ and centers $R_{I}$, and $V_{I}$
is the potential due to nucleus $I$. Note that $V_{I}$ can either
be the Coulomb potential $V_{I}(\bx)=\frac{-Z_{I}}{\vert\bx-R_{I}\vert}$
or a pseudopotential. We will explore both possibilities below. 

Meanwhile, the two-electron integrals are given by 
\begin{equation}
\mathbf{V}_{ijkl}=\int\int\eta_{i}(\bx)\eta_{j}(\bx)\frac{1}{\vert\bx-\by\vert}\eta_{k}(\by)\eta_{l}(\by)\,d\bx\,d\by.\label{eq:2electron}
\end{equation}

The structure of the quantum chemistry Hamiltonian, especially the
storage of the two-electron integrals, simplifies considerably for
certain `diagonal' basis sets, which behave like smooth delta functions
in the sense that 
\begin{equation}
\int f(\bx)\eta_{i}(\bx)d\bx=w_{i}f(\bx_{i}),\label{eq:diagprop}
\end{equation}
 where $\bx_{i}$ is the spatial center of the $i$-th basis function
and $w_{i}$ is its weight, independent of $f$. This implies that
for a sufficiently smooth function $f$ one has
\begin{equation}
f(\bx)=\sum_{i}w_{i}f(\bx_{i})\eta_{i}(\bx),\label{eq:diagprop2}
\end{equation}
 i.e., the coefficients of the expansion can simply be read off by
evaluating the function at grid points $\bx_{i}$. Diagonal basis
sets include (periodic) sinc functions\cite{boyd2001chebyshev} as well as more
recently introduced localized alternatives such as the gausslets\cite{White2017}
and their various extensions\cite{White2019,Qiu2021,White2023}.

With a diagonal basis one needs to retain only a one-index quantity
to store the nuclear integrals (instead of a two-index quantity) and
two-index quantity to store the 2-electron Coulomb integrals (instead
of a four-index quantity). In fact, as we shall explore below, it
is possible to deal with the Coulomb integrals in matrix-free fashion,
avoiding the need to even form this two-index quantity.

To see this, let us assume that the subscripts $p,q,r,s$ are used
to represent an arbitrary molecular orbital (MO). Then, abusing notation
slightly by overloading $\mathbf{U}$, the nuclear matrix in the MO
basis is given by 
\begin{align}
\mathbf{U}_{pq} & =\sum_{I}\int f_{p}(\bx)f_{q}(\bx)V_{I}(\bx)\,d\bx\nonumber \\
 & \approx\sum_{I}\int\sum_{i}w_{i}f_{p}(\bx_{i})f_{q}(\bx_{i})\eta_{i}(\bx)V_{I}(\bx)\,d\bx\nonumber \\
 & =\sum_{i} w_i^2 f_{p}(\bx_{i})f_{q}(\bx_{i})\left[  \sum_{I} \frac{1}{w_{i}} \int\eta_{i}(\bx)V_{I}(\bx)\,d\bx\right]\nonumber \\
 & =\sum_{i} f_{pi} \,  f_{qi} \,\mathbf{u}_{i},\label{eq:simplenuc}
\end{align}
where $\mathbf{u}_{i}$ is defined as suggested by the last equality and $f_{pi} := w_i f_p (\bx_i)$ is the coefficient of $f_p$ in the $\{ \eta_i \}$ basis, following~\eqref{eq:diagprop2}.
The equation (\ref{eq:simplenuc}) emphasizes that one is only required
to evaluate and store the single-index quantity $\mathbf{u}_{i}$,
which is independent of the MOs.

Similarly, the two-electron integrals in the MO basis (once again
abusing notation slightly) are given by 
\begin{align}
\mathbf{V}_{pqrs} & =\int\int f_{p}(\bx)f_{q}(\bx)\frac{1}{\vert\bx-\bx'\vert}f_{r}(\bx')f_{s}(\bx')\,d\bx\,d\bx'\nonumber \\
 & \approx\sum_{ij} w_i^2 f_{p}(\bx_{i})f_{q}(\bx_{i})\left[ \frac{1}{w_{i}w_{j}} \int\int\eta_{i}(\bx)\frac{1}{\vert\bx-\bx'\vert}\eta_{j}(\bx')\,d\bx\,d\bx'\right] w_j^2 f_{r}(\bx_{j})f_{s}(\bx_{j})\nonumber \\
 & =\sum_{ij} f_{pi}\, f_{qi} \,\mathbf{v}_{ij}\, f_{rj}  \, f_{sj}, \label{eq:2-electron}
\end{align}
i.e., the two-electron integrals in the MO basis are determined by
the two-index quantity $\mathbf{v}_{ij}$ which is independent of
the MOs. Note the similarity of (\ref{eq:2-electron}) to the tensor
hypercontraction (THC) form\cite{THC1,THC2} of the two-electron integrals
integrals. However, unlike conventional THC, used as a tool for compressing
a given GTO basis, this approach will yield complete basis set results
when the number of diagonal basis functions is taken to be sufficiently
large.

Note that the approximations (\ref{eq:simplenuc})
and (\ref{eq:2-electron}) for the nuclear and two-electron integrals
in the MO basis ($p,q,r,s$) would hold if it were the case that the nuclear and
two-electron integrals (\ref{eq:1electron}) and (\ref{eq:2electron})
in the diagonal basis $\{\eta_{i}\}$ were diagonal i.e. 
$\mathbf{U}_{ij}\approx\mathbf{u}_{i}\,\delta_{ij}$
and $\mathbf{V}_{ijkl}\approx\mathbf{v}_{ik} \, \delta_{ij} \, \delta_{kl}$,
as can be verified by substitution of (\ref{eq:diagprop2}). Such
a condition is however sufficient but not necessary, and indeed even while
such approximations for $\mathbf{U}_{ij}$ and $\mathbf{V}_{ijkl}$
do not hold elementwise to high accuracy, the approximations
(\ref{eq:simplenuc}) and (\ref{eq:2-electron}) for $\mathbf{U}_{pq}$
and $\mathbf{V}_{pqrs}$ can still hold to high accuracy due to the
smoothness of the MOs. Moreover, any computational procedure for ground
state calculations can treat the substitutions $\mathbf{U}_{ij}\leftarrow\mathbf{u}_{i} \,\delta_{ij}$
and $\mathbf{V}_{ijkl}\leftarrow\mathbf{v}_{ik} \, \delta_{ij} \, \delta_{kl}$
as if they were exact, in the sense of pseudospectral methods.

Such a `diagonal approximation'
is not variational, since the effective one- and two-electron integrals
used in such a calculation are not obtained by exact Galerkin truncation.
Thus in particular the Hartree-Fock results presented in this work
do not converge \emph{from above} to the exact (i.e., complete basis
set) Hartree-Fock results. However, the non-variational error incurred
due to the diagonal approximation is of similar magnitude to the error
due to truncation error of the underlying diagonal basis set (for example, see Figures 8 and 9 in Ref\cite{White2017}).
At the same time, the energy obtained from these basis sets (treated
either variationally or within the diagonal approximation) converges
to the exact result only algebraically, rather than exponentially,
due to the non-analyticity of the Coulomb potential. To overcome this
slow convergence, we will make use of the all-electron pseudopotential
(PP) proposed recently by Gygi\cite{Gygi2023}. This PP can be made arbitrarily
accurate by modifying just a single parameter (called $a$ in the
original paper). For the calculations presented here we use $a=4$
which is expected to give results that are within a few tens of $\mu$E$_{h}$
of the exact results for elements in the first two rows of the periodic table. 

In this paper we show how to extend existing
grid-based diagonal basis sets by composing with an arbitrary invertible
transformation. Specifically, we will focus on the case where the
original diagonal basis set is a set of periodic sinc functions, or
sinc functions for short, whose span is identical to that of a suitable
set of planewaves adapted to a periodic box. For periodic sinc functions
the weights satisfy $w_{i}=w=\sqrt{\frac{\mathcal{V}}{M}}$, where
$\mathcal{V}$ is the volume of the computational box and $M$ is
the number of basis functions or grid points in the domain, so $w_{i}^{2}$
can be viewed as a quadrature weight for integrals performed over
the computational grid.
Note that sinc functions have the property that $\phi_{i}(\bx_{j})=w^{-1}\delta_{ij}$.
Moreover, for sinc functions it is well-known that (\ref{eq:diagprop})
is satisfied for suitably band-limited functions. Further details
on this choice of basis are given in Appendix~\ref{sec:sincdefs}.

\section{Diagonal approximation with the deformed basis} \label{sec:diagonaldef}

One of the shortcomings of sinc and Gausslet basis is that it provides uniform resolution in the entire
domain of the problem. Thus even though high resolution is needed
only near the nucleus, one has to uniformly increase the number of
basis function over the entire computational domain to get accurate
results. This is wasteful, and it was noted is the early 1990s (by
Gygi\citep{Gygi1992,Gygi1993,Gygi1995,Gygi1995a}, followed by others\citep{Devenyi1994,JoseM.Perez-Jorda1995,Perez-Jorda1998,Perez-Jorda2011,Perez-Jorda2014,Hamann1995,Hamann1996,Hamann2001,Hamman1995,Modine1997,Zumbach1996,Rodriguez2008}) that one can overcome
this shortcoming by introducing a deformation that preferentially
introduces a larger number of basis functions or grid points near the nucleus.

To see how this works, let us say we are given a collection of sinc
functions $\{\phi_{i}\}$ centered at uniform rectangular grid points
$\{\by_{i}\}$. Let us now introduce an invertible transformation
$T:\R^{3}\ra\R^{3}$. Then a deformed diagonal basis set $\{\eta_{i}\}$
with centers $\{\bx_{i}=T^{-1}(\by_{i})\}$ can be constructed formally
as 
\begin{align*}
\eta_{i}(\bx) & :=\psi_{i}(\bx)\sqrt{J(\bx)},\\
\psi_{i}(\bx) & :=\phi_{i}(T(\bx))
\end{align*}
 where $J(\bx):=\det DT(\bx)$ is the Jacobian determinant of the
deformation map and can be interpreted as the specifying the density
of point at position $\bx$. In the following it is useful to set
the following notation: 
\begin{itemize}
\item $\phi_{i}$ denote the sinc basis functions.
\item $\by_{i}$ denote the uniform grid of sinc centers.
\item $\bx_{i}=T^{-1}(\by_{i})$ denote the deformed grid points.
\item $\psi_{i}(\bx):=\phi_{i}(T(\bx))$ denote the transformed sinc functions, which inherit the interpolating property.
\item $\eta_{i}(\bx):=\phi_{i}(T(\bx))\,\sqrt{J(\bx)}=\psi_{i}(\bx)\,\sqrt{J(\bx)}$
denote the orthogonal diagonal basis functions.
\item $J_{i}:=J(\bx_{i})$ denote the determinant of the Jacobian at the
deformed grid points.
\end{itemize}
It is important to note the distinction between $\psi_{i}$ and $\eta_{i}$.
The latter form the orthogonal diagonal basis that we actually use
as the quantum chemistry basis set. However, for a `suitably smooth'
$f$ we can calculate 
\[
f(T^{-1}(\by))\approx w\sum_{j}f(T^{-1}(\by_{j})) \, \phi_{j}(\by)
\]
 by evaluating (\ref{eq:diagprop2}) at $T^{-1}(\by)$, which implies
that for arbitrary $\bx$, 
\begin{align}
f(\bx) & \approx w\sum_{j}f(\bx_{j}) \, \psi_{j}(\bx).\label{eq:psiinterp}
\end{align}
 Therefore the functions $\psi_{i}$ (not $\eta_{i}$) inherit the
interpolating property (\ref{eq:diagprop2}) from the sinc functions,
as well as the property that $\psi_{i}(\bx_{j})=w^{-1}\delta_{ij}$.

In the above discussion, the precise meaning of `suitably smooth'
is that $f\circ T^{-1}$ is sufficiently bandlimited. Therefore the
$\psi_{i}$ functions are good interpolators even for functions with
sharp features as long as the the density $J(\bx)$ of points is also
large in these regions---and moreover with only dull features where
$J(\bx)$ is small.

Evidently we can also approximate 
\begin{equation}
f(\bx)\approx w\sum_{j}\frac{f(\bx_{j})}{\sqrt{J_{j}}}\eta_{j}(\bx),\label{eq:etaInterp}
\end{equation}
 as can be obtained by using the interpolating basis $\{\psi_{j}\}$
to expand $f(\bx)/\sqrt{J(\bx)}$ instead of $f(\bx)$. For electronic
structure calculations we note that the basis set $\{\eta_{j}\}$
is convenient because it is orthogonal. For mean-field calculations
we often go between representations in the $\{\eta_{j}\}$ and $\{\psi_{j}\}$
basis representations given in (\ref{eq:etaInterp}) and (\ref{eq:psiinterp}),
respectively, by simply multiplying the coefficients in the $\{\eta_{j}\}$
expansion by $\sqrt{J_{j}}$ to get the coefficients in the $\{\psi_{j}\}$
expansion. The expansion in terms of $\{\psi_{j}\}$ is more convenient
when one wants to apply the kinetic and Coulomb operators (cf. Section
\ref{sec:pseudo} below), while expansion in terms of $\{\eta_{j}\}$
is more convenient for calculating overlaps between orbitals and maintaining
orthogonality.

Using (\ref{eq:etaInterp}) it can
be readily shown that the nuclear integrals of MOs will be given (up
to the approximation introduced above) by
\begin{equation}
\label{eq:boldu}
\mathbf{U}_{pq} =\sum_{i} w^2 \frac{f_{p}(\bx_{i})}{\sqrt{J_{i}}}\frac{f_{q}(\bx_{i})}{\sqrt{J_{i}}}\left[ \frac{\sqrt{J_{i}}}{w} \sum_{I}\int\eta_{i}(\bx)V_{I}(\bx)\,d\bx\right] =\sum_{i} f_{pi} \,  f_{qi} \, \mathbf{u}_{i}, 
\end{equation}
where $\mathbf{u}_{i}$ is the bracketed quantity in the second expression,
adapting the definition in (\ref{eq:simplenuc}) to the deformed basis. Meanwhile $f_{pi} = f_p (\bx_i) / \sqrt{J_i}$ are once again the coordinates of the MO $f_p$ in the orthonormal $\{ \eta_i \}$ basis.

Similarly the 2-electron integrals can be written as
\begin{align}
\mathbf{V}_{pqrs} & =\sum_{ij} w^2 \frac{f_{p}(\bx_{i})}{\sqrt{J_{i}}}\frac{f_{q}(\bx_{i})}{\sqrt{J_{i}}}\left[  \frac{\sqrt{J_i \, J_j} }{w^2} \int\int\eta_{i}(\bx)\frac{1}{|\bx-\bx'|}\eta_{j}(\bx')\,d\bx\,d\bx' \right] w^2 \frac{f_{r}(\bx_{j})}{\sqrt{J_{j}}}\frac{f_{s}(\bx_{j})}{\sqrt{J_{j}}} \nonumber \\
 & =\sum_{ij} f_{pi} \, f_{qi} \, \mathbf{v}_{ij} \, f_{rj} \, f_{sj} \label{eq:boldv}
\end{align}
where the two-index quantity $\mathbf{v}_{ij}$, adapting (\ref{eq:2-electron}),
is given by the bracketed quantity in the first line.

Alternatively, by instead using \eqref{eq:psiinterp} we can similarly derive 
\begin{equation*}
    \mathbf{U}_{pq} = \sum_{i} f_{pi} \, f_{qi} \, \tilde{\mathbf{u}}_i,
    \qquad
    \mathbf{V}_{pqrs} = \sum_{ij} f_{pi} \, f_{qi} \, \tilde{\mathbf{v}}_{ij} \, f_{rj} \, f_{sj},
\end{equation*}
where 
\begin{equation}
\label{eq:uvtilde}
    \tilde{\mathbf{u}}_i :=
    \frac{{J_{i}}}{w} \sum_{I}\int\psi_{i}(\bx)V_{I}(\bx)\,d\bx,
    \qquad
    \tilde{\mathbf{v}}_{ij} := 
     \frac{{J_i \, J_j} }{w^2} \int\int\psi_{i}(\bx)\frac{1}{|\bx-\bx'|}\psi_{j}(\bx')\,d\bx\,d\bx'.
\end{equation}


Similarly to the original sinc basis, the deformed sinc basis, together
with the suitable diagonal approximation, allows one to reduce the
two-electron integrals from a 4-index quantity to a 2-index quantity. Once again, to construct an electronic structure Hamiltonian~\eqref{eq:secondquant} with accurate ground-state properties, we can perform the effective substitutions $\mathbf{U}_{ij}\leftarrow\mathbf{u}_{i} \,\delta_{ij}$
and $\mathbf{V}_{ijkl}\leftarrow\mathbf{v}_{ik} \, \delta_{ij} \, \delta_{kl}$, or alternatively  
$\mathbf{U}_{ij}\leftarrow\tilde{\mathbf{u}}_{i} \,\delta_{ij}$
and $\mathbf{V}_{ijkl}\leftarrow\tilde{\mathbf{v}}_{ik} \, \delta_{ij} \, \delta_{kl}$, within the one- and two-electron integrals, in place of the exact Galerkin formulations of \eqref{eq:1electron} and \eqref{eq:2electron}.  
However, compared to the sinc basis (as we shall verify in the numerical studies of 
Section~\ref{sec:numerics}),
significantly fewer deformed sinc basis functions are needed for high
accuracy results, given an appropriately chosen deformation $T$.

Unfortunately, the deformed sinc basis functions do not enjoy the
same separable form as the original sinc basis functions, if the deformation
$T$ is not of separable form. Thus the integrals $\mathbf{u}_{i},\mathbf{v}_{ij}$, as well as the suitable matrix elements $\mathbf{T}_{ij}$ of the kinetic operator defined in \eqref{eq:1electron}, are nontrivial to evaluate for a general deformation $T$. Furthermore,
although it is possible to specify a density $J(\bx)$ of deformed
grid points, it is not trivial to specify a map $T$ that is consistent
with the prescribed density.

In the following two sections, we will explain how to approach these
two key computational difficulties of working with such a basis. First,
in Section \ref{sec:transformation}, we will show how to efficiently
construct a map $T(\bx)$ achieving a prescribed Jacobian $J(\bx)$,
i.e., a prescribed density of deformed basis function centers. Second,
in Section \ref{sec:pseudo}, we will show how the computation of
the integrals $\mathbf{u}_{i},\mathbf{v}_{ij},\mathbf{T}_{ij}$ can
be simplified with a pseudospectral approach. In fact we will show
that matrix-vector multiplication by suitable matrix quantities can be performed efficiently without forming them as dense matrices, which will imply that for mean-field calculations
all operations can be performed at a cost that is log-linear in the
number of basis functions instead of quadratic. Thus we shall achieve
the same asymptotic scaling with respect to the number of basis function
as planewave calculations, but with a smaller basis set.

\section{Constructing the coordinate transformation \label{sec:transformation}}

Suppose we are given an arbitrary probability density $\rho$, which
is not to be conflated with the electron density, on a box $\mathcal{B}:=\prod_{i=1}^{d}[a_{i},b_{i}]$
of volume $\mathcal{V}$ in $\R^{d}$. We want to find a map $T:\mathcal{B}\ra\mathcal{B}$
such that the Monge-Amp\`ere equation~\cite{Villani_2003} 
\begin{equation}
\det(DT(\bx))=\mathcal{V}\,\rho(\bx)\label{eq:mae}
\end{equation}
 holds for $x\in\mathcal{B}$. The solution map $T$ pushes forward
the density $\rho$ on $\mathcal{B}$ to the uniform density $\mathcal{V}^{-1}$
on $\mathcal{B}$. Note that the right-hand side is the density ratio
of $\rho(\bx)$ to the target density at $T(\bx)$, which is simply
the uniform density $\mathcal{V}^{-1}$ on the box. It is useful to
define $L_{i}:=b_{i}-a_{i}$ to be the length of the box in each dimension,
so $\mathcal{V}=\prod_{i=1}^{d}L_{i}$.

In the context of the deformed basis
construction outlined above, our goal is to solve (\ref{eq:mae})
where $\rho(\bx)\propto J(\bx)$ and $J$ is a prescribed density
of deformed basis function centers. However, we will consider (\ref{eq:mae})
more abstractly because ultimately our approach will view the solution
of (\ref{eq:mae}), for arbitrary right-hand side $\rho(\bx)$, as
a computational primitive that we shall wrap in several layers of
abstraction in order to obtain a deformation of high quality, as efficiently
as possible.

We will build our approach in several stages. In Section \ref{sec:reviewKR}
we review an essential ingredient used to build our map: the Knothe-Rosenblatt
transport~\cite{Rosenblatt,Knothe}. In Section \ref{sec:periodic}, we explain how
the Knothe-Rosenblatt transport extends to the setting of a periodic
computational domain. In Section \ref{sec:separableKR}, we show how
the Knothe-Rosenblatt transport can be computed exactly when $\rho$
is presented as a sum of separable functions. In Section \ref{sec:cyclicKR},
we explain how several Knothe-Rosenblatt transports (with the roles
of the variables cyclically permuted) can be composed to produce solutions
to the Monge-Amp\`ere equation (\ref{eq:mae}) which are preferable
to the unmodified Knothe-Rosenblatt transport. In particular we explain
how to fit each transport problem in the sequence in the form (\ref{eq:mae})
with separable right-hand side $\rho$. We call this approach the
cyclic Knothe-Rosenblatt flow.

Finally, in Section \ref{sec:finalKR}, we wrap the cyclic Knothe-Rosenblatt
flow in one more layer of abstraction. This final complication is
motivated by the fact that, as we shall explain, it is computationally
preferable to compute the inverse map $S(\by)=T^{-1}(\by)$, which
necessarily satisfies the `inverse' Monge-Amp\`ere equation 
\begin{equation}
\det(DS(\by))\propto\frac{1}{J(S(\by))},\label{eq:invmae}
\end{equation}
 where $J$ is the prescribed density of basis function centers. This
formulation is more difficult to solve directly due to the dependence
of the right-hand side on the unknown map $S$ itself. However, motivated
by the approach of Lindsey and Rubinstein~\cite{LindseyRubinstein2017}, we will show how (\ref{eq:invmae})
can be solved by computing a convergence sequence of solutions of
problems of the form (\ref{eq:mae}), where $\rho$ is determined
self-consistently.

\subsection{Review of Knothe-Rosenblatt transport \label{sec:reviewKR}}

The Monge-Amp\`ere equation is underdetermined. Many solutions can
be constructed as optimal transport maps~\cite{Villani_2003}, but there is
in fact a semi-explicit solution that forms the building block of
our approach. To wit, the Knothe-Rosenblatt transport~\cite{Rosenblatt,Knothe} is
the unique solution of (\ref{eq:mae}) among `triangular maps' i.e.,
maps of the form $T=(T_{1},\ldots,T_{d})$, where each $T_{k}$ depends
only on $x_{1},\ldots,x_{k}$. Thus, e.g., for $d=3$, 
\[
DT(\bx)=\left(\begin{array}{ccc}
\frac{\partial T_{1}}{\partial x_{1}}(x_{1}) & 0 & 0\\
* & \frac{\partial T_{2}}{\partial x_{2}}(x_{1},x_{2}) & 0\\
* & * & \frac{\partial T_{3}}{\partial x_{3}}(x_{1},x_{2},x_{3})
\end{array}\right),
\]
 and $\det(DT)$ is the product of the diagonal terms. This solution was used previous by Perez-Jorda\cite{JoseM.Perez-Jorda1995} to solve the Monge-Amp\`ere equation in the context of solution of electronic structure with adaptive coordinates.

Let $\rho_{1}(x_{1})$ be the first marginal distribution of $\rho$,
and $\rho_{2}(x_{1},x_{2})$, etc. denote the marginals for the first
several variables. Moreover, let $\rho_{2}(x_{2}\vert x_{1})=\rho_{2}(x_{1},x_{2})/\rho_{1}(x_{1})$
and $\rho_{3}(x_{3}\vert x_{1},x_{2})=\rho(x_{1},x_{2},x_{3})/\rho(x_{1},x_{2})$,
etc. denote the conditional densities.

It follows from the Monge-Amp\`ere equation (\ref{eq:mae}) and our
restriction on the form of $T$ that the equations 
\[
\frac{\partial T_{1}}{\partial x_{1}}(x_{1})=L_{1}\,\rho_{1}(x_{1}),\quad\frac{\partial T_{2}}{\partial x_{2}}(x_{1},x_{2})=L_{2}\,\rho_{2}(x_{2}\vert x_{1}),\quad\frac{\partial T_{3}}{\partial x_{3}}(x_{1},x_{2},x_{3})=L_{3}\,\rho_{3}(x_{3}\vert x_{1},x_{2}),
\]
 etc., must hold.

The first equation is solved simply with an antiderivative: 
\begin{equation}
T_{1}(x_{1})=a_{1}+L_{1}\int_{a_{1}}^{x_{1}}\rho_{1}(y_{1})\,dy_{1}.\label{eq:T1}
\end{equation}
 Evidently, $T_{1}$ satisfies the boundary conditions $T_{1}(a_{1})=a_{1}$
and $T_{1}(b_{1})=b_{1}$.

Next we solve the second equation: 
\begin{equation}
T_{2}(x_{1},x_{2})=a_{2}+L_{2}\int_{a_{2}}^{x_{2}}\rho_{2}(y_{2}\vert x_{1})\,dy_{2},\label{eq:T2}
\end{equation}
 which also achieves the second boundary condition $T_{2}(x_{1},a_{2})=a_{2}$
and $T_{2}(x_{1},b_{2})=b_{2}$.

Then the third equation: 
\begin{equation}
T_{3}(x_{1},x_{2},x_{3})=a_{3}+L_{3}\int_{a_{3}}^{x_{3}}\rho_{2}(y_{3}\vert x_{1},x_{2})\,dy_{2},\label{eq:T3}
\end{equation}
and so on.

It follows from the construction that 
\[
T_{i}(x)=\begin{cases}
a_{i} & \text{if}\ x_{i}=a_{i}\\
b_{i} & \text{if}\ x_{i}=b_{i},
\end{cases}
\]
 i.e., $T$ fixes the boundary of $\mathcal{B}$.

It may be useful to compute the inverse map $S=T^{-1}$. In fact $S=(S_{1},\ldots,S_{d})$
has the same form in that $S_{1}=S_{1}(y_{1})$, $S_{2}=S_{2}(y_{1},y_{2})$,
etc. Evidently $S_{1}(y_{1})=T_{1}^{-1}(y_{1})$ can be obtained by
univariate inversion of $T_{1}$. Once we have found $x_{1}=S_{1}(y_{1})$,
we want to find $x_{2}$ such that 
\[
T_{2}(x_{1},x_{2})=y_{2},
\]
 which is achieved by univariate inversion of the function $T_{2}(x_{1},\,\cdot\,)$.
The result $x_{2}$ is $S_{2}(y_{1},y_{2})$. Similarly, once we have
found $x_{1},x_{2}$, we can obtain $x_{3}=S_{3}(y_{1},y_{2},y_{3})$
by inverting the function $T_{2}(x_{1},x_{2},\,\cdot\,)$.

\subsection{Periodic extension \label{sec:periodic}}

In the case that $\rho$ extends smoothly and periodically from the
box $\mathcal{B}$ to $\R^{d}$, in fact the map $T:\mathcal{B}\ra\mathcal{B}$
also extends smoothly and periodically, in the sense that the coordinate
functions of $T-\mathrm{Id}$ are all periodic, and the extension
preserves each unit cell. We prove this claim in Appendix \ref{sec:periodicboundary}.

\subsection{Explicit solution for sum of separable functions \label{sec:separableKR}}

Suppose $\rho$ has the functional form 

\[
\rho(\bx)=\sum_{\alpha}c^{\alpha}g^{\alpha}(\bx),
\]
 where each term $g^{\alpha}(\bx)=\prod_{i=1}^{d}g_{i}^{\alpha}(x_{i})$
is separable. Let $G_{i}^{\alpha}(x_{i})=\int_{a_{i}}^{x_{i}}g_{i}^{\alpha}(y_{i})\,dy_{i}$
be the appropriate antiderivative, and assume that we can evaluate
$g_{i}^{\alpha},G_{i}^{\alpha}$ easily. We show in Appendix \ref{sec:separable}
how the exact Knothe-Rosenblatt transport can be computed using such
evaluations.

It is convenient to think of the basis index $\alpha$ as a multi-index
$\alpha=(n_{1},\ldots,n_{d})$ and define 
\[
g_{i}^{\alpha}=P_{n_{i}}\circ u_{i},
\]
 where $P_{n}$ is the $n$-th Chebyshev polynomial, defined on the
standard reference domain $[-1,1]$, and $u_{i}$ is the linear map
sending $[a_{i},b_{i}]$ to $[-1,1]$. This choice defines an efficient
basis for the expansion of the density $\rho$, and in Appendix \ref{sec:chebyshev}
we explain how the exact computation of Appendix \ref{sec:separable}
can be carried out explicitly in this setting. This discussion amount
to the computation of $G_{i}^{\alpha}$ in the special case at hand.

In the case of periodic $\rho$, we can likewise expand in a basis
$g_{k}^{\alpha}=e_{n_{k}}\circ u_{k}$, where $e_{n}(x)=e^{2\pi in_{k}x}$
is a suitable Fourier mode and $u_{k}$ maps $[a_{k},b_{k}]$ linearly
to $[0,1]$. In this case the computation of $G_{k}^{\alpha}$ is
even simpler, cf. Appendix \ref{sec:fourier}.

\subsection{Cyclic Knothe-Rosenblatt flow \label{sec:cyclicKR}}

\begin{figure}
\noindent\begin{centering}\includegraphics[width=0.5\textwidth, trim={0 50 0 50},clip]{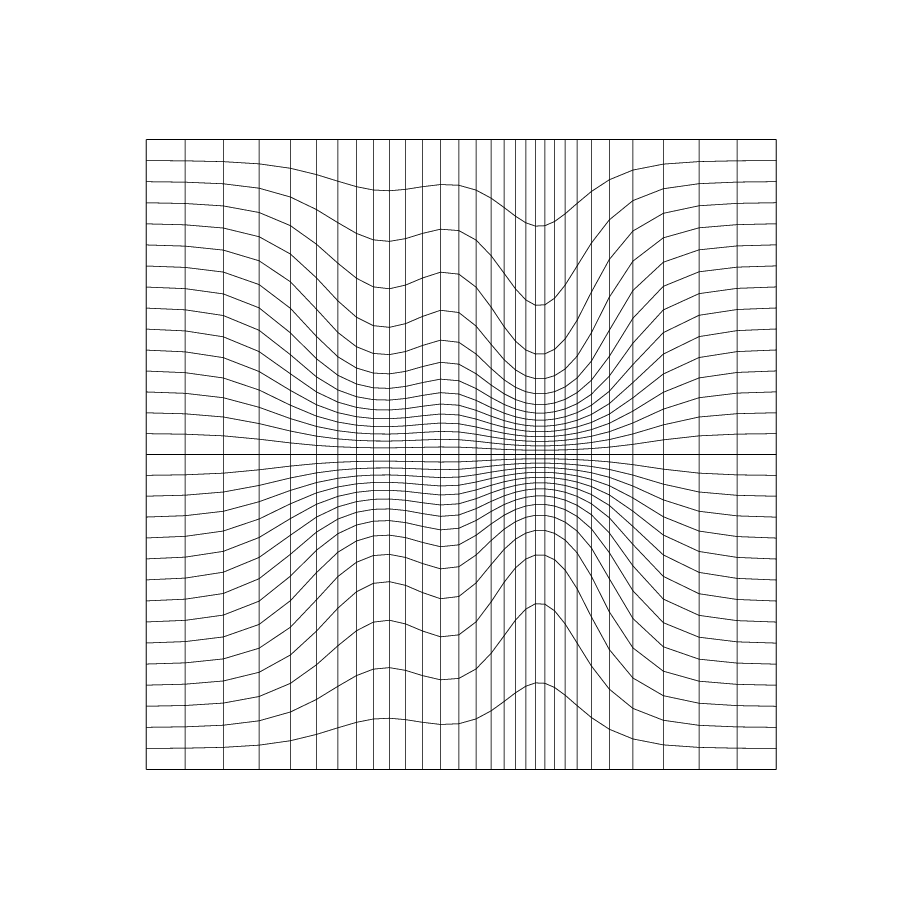}\includegraphics[width=0.5\textwidth, trim={0 50 0 50},clip]{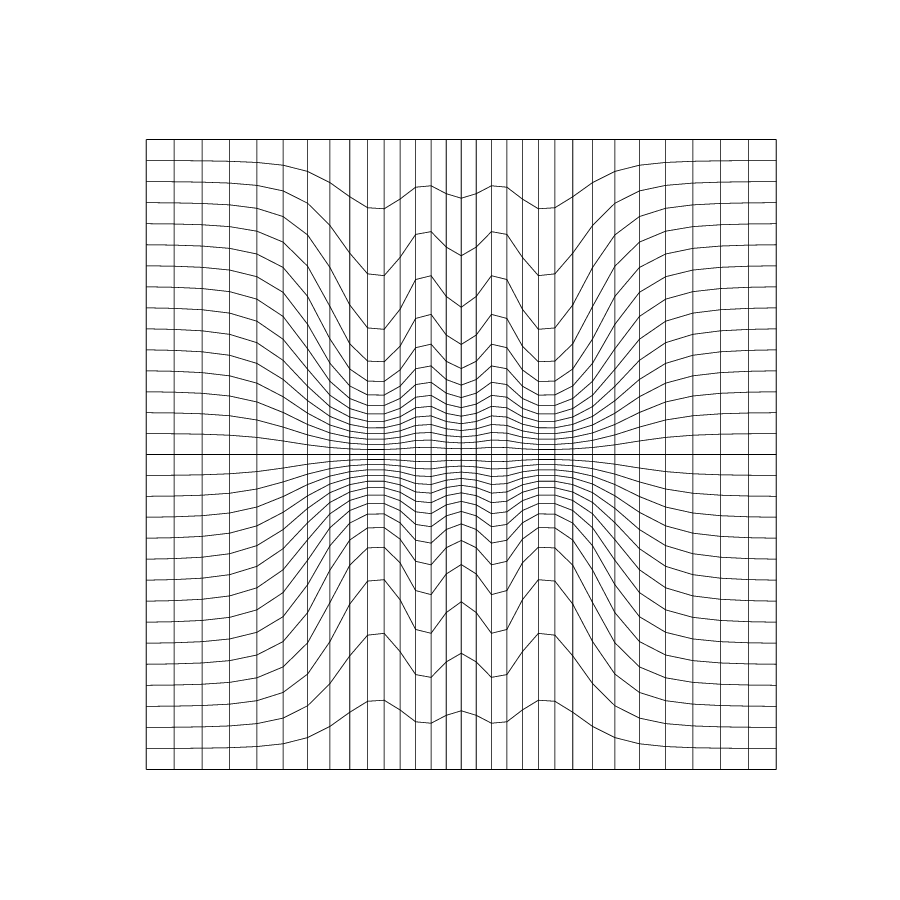}
\end{centering}
\caption{Consider $\rho (\bx) \propto 0.01 + \sum_{I=1}^3 e^{-3 \sqrt{ 0.3 + \vert \bx - R_I\vert^2 } }$ on the non-periodic 3-dimensional box $[-4,4]^3$, where $R_1 = (0,1,0)$, $R_2 = (1,-1,0)$, and $R_3 = (-1,-1,0)$. We plot the curvilinear coordinates $\bx = T^{-1} (\by)$ furnished by naive Knothe-Rosenblatt transport. Left: image of $y_2$-$y_3$ coordinate lines in $x_2$-$x_3$ plane for fixed $y_1 = 0$. Right: image of $y_1$-$y_3$ coordinate lines in $x_1$-$x_3$ plane for fixed $y_2 = 0$. Observe that vertical lines are preserved by the coordinate transformation.}
\label{fig:awk}
\end{figure}

A single Knothe-Rosenblatt transport is precisely the type of category
of map considered in the context of the multi-sliced Gausslet basis~\cite{White2019}, which
was applied to quasi-1D molecular geometries. However, the Knothe-Rosenblatt
transport treats the coordinates $x_{1},x_{2},x_{3}$ asymetrically
and can produce awkward deformations in more general cases where it
is not possible to define a geometrically natural ordering of the
coordinates, as illustrated in Figure \ref{fig:awk}.

As such it is desirable to restore the permutation symmetry among
the coordinates. We will achieve this goal by incrementally building
a transport map as a composition of small Knothe-Rosenblatt transports
in which the roles of the coordinates are cycled. We call the resulting
map the \emph{cyclic Knothe-Rosenblatt flow}.

Concretely, we will build 
\[
T=T^{(N)}\circ\cdots\circ T^{(1)},
\]
 where we choose $T^{(1)}$ such that 
\[
\det(DT^{(1)}(\bx))\propto\rho^{1/N}(\bx)
\]
 for all $\bx$.

Then we choose $T^{(2)}$ such that 
\[
\det(DT^{(2)}(\bx))\propto\rho^{1/N}\left([T^{(1)}]^{-1}(\bx)\right),
\]
 or equivalently 
\[
\det(DT^{(2)}(T^{(1)}(\bx)))\propto\rho^{1/N}\left(\bx\right),
\]
 for all $\bx$.

This ensures that 
\begin{align*}
\det\left(D\left[T^{(2)}\circ T^{(1)}\right](\bx)\right) & =\det\left(DT^{(2)}(T^{(1)}(\bx))\,DT^{(1)}(\bx)\right)\\
 & =\det\left(DT^{(2)}(T^{(1)}(\bx))\right)\det\left(DT^{(1)}(\bx)\right)\\
 & \propto\rho^{1/N}(x)\,\rho^{1/N}(\bx)\\
 & =\rho^{2/N}(\bx).
\end{align*}

More generally, we construct $T^{(n)}$ inductively in terms of $T^{(n-1)}$
to ensure that
\[
\det(DT^{(n)})\propto \rho^{(n)},
\]
 where
\[
\rho^{(n)}\propto\rho^{1/N}\circ[T^{(n-1)}\circ\cdots\circ T^{(1)}]^{-1}.
\]

Once $\rho^{(n)}$ is fit with a separable expansion (e.g., Chebyshev
or Fourier), the map $T^{(n)}$ can be computed according to the procedure
in Section \ref{sec:separableKR} for computing the Knothe-Rosenblatt
transport. However, as we iterate $n=1,\ldots,N$, we cycle the roles
of the coordinates before computing the Knothe-Rosenblatt transport.

How do we fit each $\rho^{(n)}$ with a separable expansion? It is useful
to note that 
\[
\rho^{(n)}\propto \rho^{(n-1)}\circ\left[T^{(n-1)}\right]^{-1}.
\]
 Therefore we can evaluate $\rho^{(n)}$ on Chebyshev nodes provided
we can invert the incremental map $T^{(n-1)}$ and evaluate the previous
right-hand side $\rho^{(n-1)}$, which we assume inductively is already
fit with a Chebyshev expansion. Once we have evaluated $\rho^{(n)}$
on the Chebyshev nodes, the fitting with Chebyshev polynomials follows
from standard techniques. The same approach applies \emph{mutatis
mutandi} to the periodic case where the right-hand sides may be fit
with Fourier expansions by evaluation on an equispaced grid.

\begin{figure}
\noindent\begin{centering}\includegraphics[width=0.5\textwidth, trim={0 50 0 50},clip]{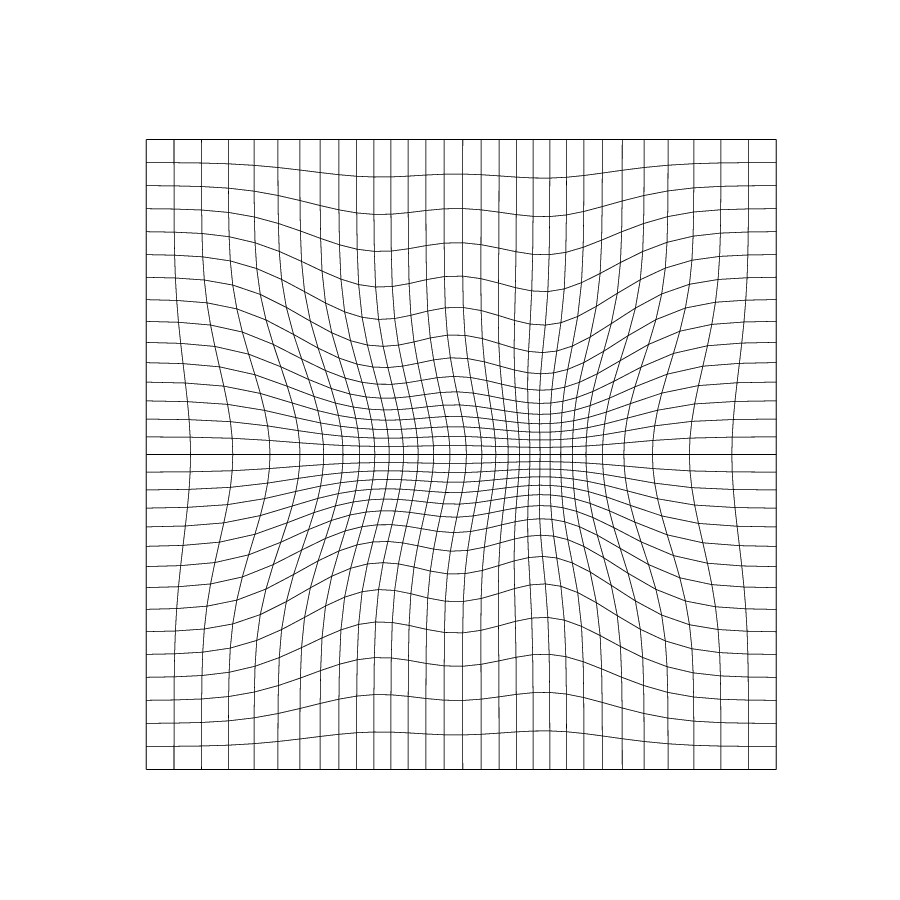}\includegraphics[width=0.5\textwidth, trim={0 50 0 50},clip]{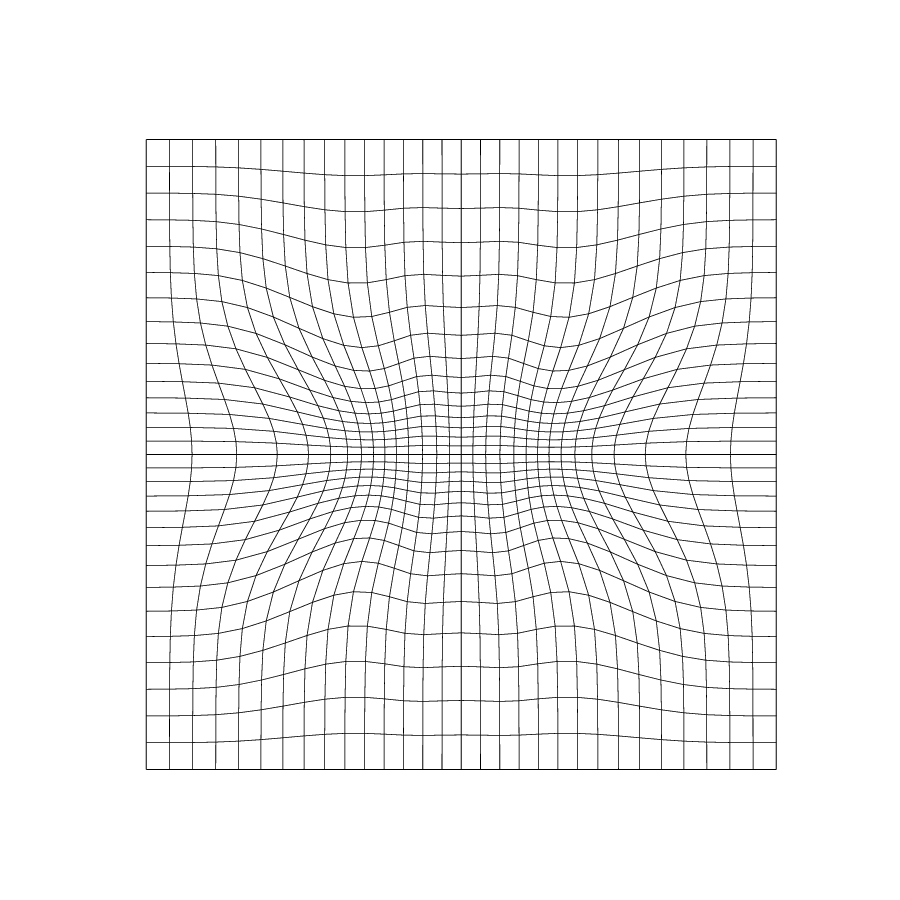}
\end{centering}
\caption{Just as in Figure~\ref{fig:awk}, we consider $\rho (\bx) \propto 0.01 + \sum_{I=1}^3 e^{-3 \sqrt{ 0.3 + \vert \bx - R_I\vert^2 } }$ on the non-periodic 3-dimensional box $[-4,4]^3$, where $R_1 = (0,1,0)$, $R_2 = (1,-1,0)$, and $R_3 = (-1,-1,0)$. We plot the curvilinear coordinates $\bx = T^{-1} (\by)$ furnished by cyclic Knothe-Rosenblatt flow with $N=40$. Left: image of $y_2$-$y_3$ coordinate lines in $x_2$-$x_3$ plane for fixed $y_1 = 0$. Right: image of $y_1$-$y_3$ coordinate lines in $x_1$-$x_3$ plane for fixed $y_2 = 0$.}
\label{fig:nonawk}
\end{figure}

The cyclic Knothe-Rosenblatt flow is illustrated in Figure \ref{fig:nonawk} using the same choice of $\rho$ as in our demonstration of the ordinary Knothe-Rosenblatt transport in Figure~\ref{fig:awk}. Compared to the ordinary Knothe-Rosenblatt transport, our flow yields a smoother transformation with significantly better-conditioned coordinate boxes.

Once the cyclic Knothe-Rosenblatt flow $T$ is constructed, $T^{-1}(\by)$
can be evaluated by successively inverting the incremental maps $T^{(n)}$.
Since these are all triangular maps, they can be inverted in three
space dimensions by solving $d=3$ successive univariate equations.
However, since each incremental map is a small perturbation of the
identity, it can also be inverted efficiently be a few iterations
of fixed-point iteration. Concretely, if we want to solve $\by=T^{(n)}(\bx)$
for $\bx$, we can look for a fixed point of the map $\bx\mapsto\bx-\left[T^{(n)}(\bx)-\by\right]$
using fixed-point iteration (possibly with mixing to aid convergence).

We comment that the evaluation of a single layer of Knothe-Rosenblatt transport is easier to perform on a separable grid than an arbitrary collection of points, as can be readily observed from the conclusion of Appendix \ref{sec:fourier}. However, several steps in our approach require evaluation of such a layer on a non-separable collection of points. Specifically, the evaluation of the inverse maps $T^{(n)}$ by fixed-point iteration requires evaluations on unstructured grids. Moreover, a full forward pass through the transport as a composition of the incremental maps $T^{(n)}$ necessarily requires evaluation on unstructured grids. In addition, the fitting step wherein $\rho^{(n)}$ is evaluated on a separable grid requires us to evaluate the previous $\rho^{(n-1)}$ on an unstructured grid, which is similarly inefficient. We accelerate all of these steps using non-uniform fast Fourier transforms (NUFFTs)~\cite{BOYD1992243,DuttRokhlin} to perform fast and highly accurate interpolation of such unstructured evaluations from a fast evaluation on a separable grid, maintaining log-linear scaling in the size of our computational grid throughout the implementation. We use the FiNUFFT library \cite{barnett_parallel_2019,cufinufft,barnett_aliasing_2021} to evaluate our NUFFTs. 

\subsection{Self-consistent inverse transport \label{sec:finalKR}}

As mentioned in the opening of Section \ref{sec:transformation},
we wrap the cyclic Knothe-Rosenblatt transport in one final computational
layer.

Our motivation is as follows. Suppose we try to solve the Monge-Amp\`ere
equation (\ref{eq:mae}), where the right-hand side is given by $\rho(\bx)\propto J(\bx)$.
Within each layer of the computation of the cyclic Knothe-Rosenblatt
transport, we must fit the right-hand side $\rho^{(n)}$ with a sum of
separable functions, by pointwise evaluation on a separable computational
grid. However, $\rho(\bx)$, and therefore the intermediate functions
$\rho^{(n)}(\bx)$ as well, are very sharp precisely in the case where
$J(\bx)$ is sharp. Just as we want to avoid resolving the wavefunction
with uniform resolution across the entire computational domain, so
also we want to avoid resolving the nonuniform density $J(\bx)$ of
basis functions with uniform accuracy across the entire computational
domain.

\begin{figure}
\noindent\begin{centering}\includegraphics[width=0.5\textwidth, trim={0 50 0 50},clip]{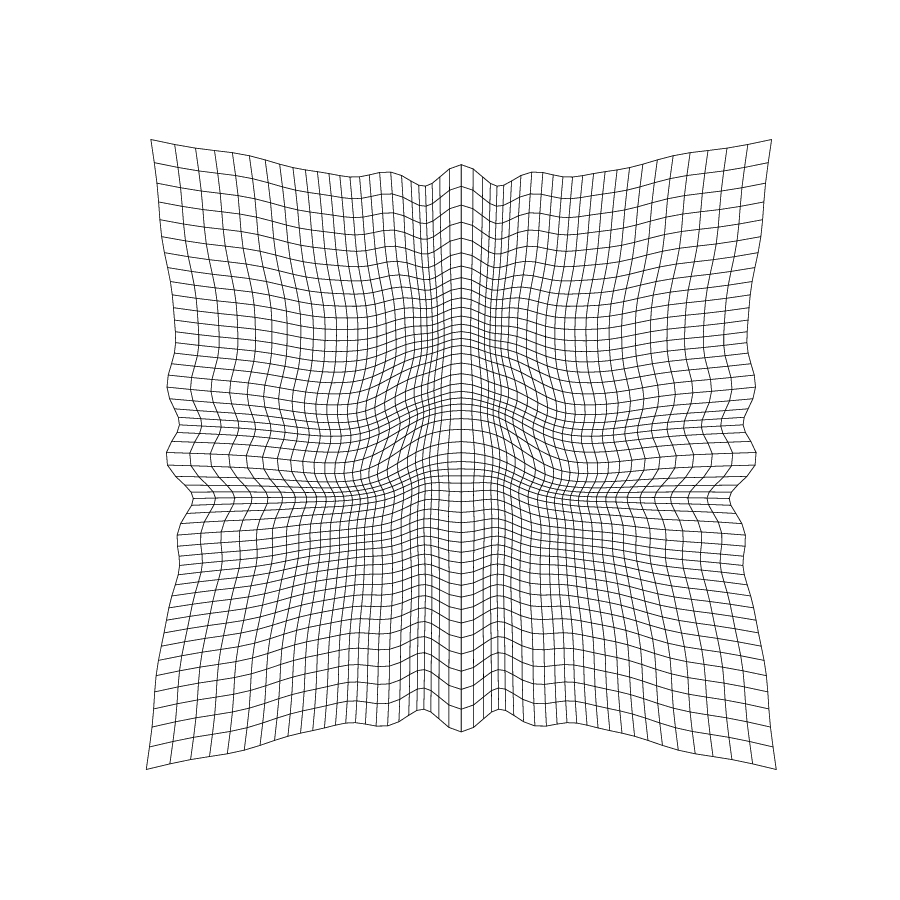}\includegraphics[width=0.5\textwidth, trim={0 50 0 50},clip]{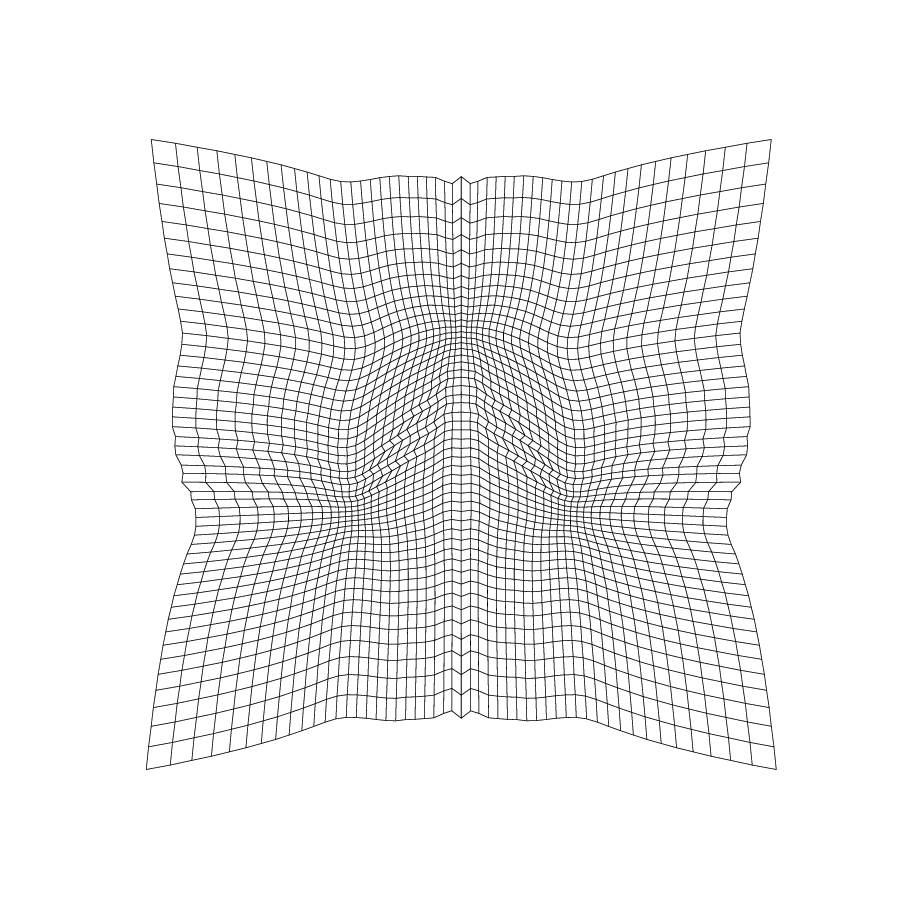}
\end{centering}
\caption{Image of the subset $[-4,4]^2 \times \{0\}$ of the computational domain $[-5,5]^3$ under $T^{-1}$, where $T$ is computed via naive cyclic Knothe-Rosenblatt flow and $\rho$ is specified in the main text. Left: result using 21 computational grid points per dimension. Right: result using 65 computational grid points per dimension. Observe that significant aliasing error is still present even for the finer computational grid.}
\label{fig:alias}
\end{figure}

The difficulty is illustrated in Figure~\ref{fig:alias}, which illustrates results for $\rho (\bx)$ chosen following the functional form \eqref{eq:deform} that we use in our experiments below, where we take $a=0.1$, $b=4$, $c=0.01$, and we consider three atom centers $I=1,2,3$ with $Z_I = 1$, positioned at the vertices of an equilateral triangle on the unit circle in the $\{x_3 = 0\}$ plane. The computational box is $[-5,5]^3$ with periodic boundary conditions. The prescribed $\rho$ is very sharp, with over 3 orders of magnitude of variation in the density over the computational domain, and as a result it is too expensive to overcome aliasing error using the naive cyclic Knothe-Rosenblatt transport of Section~\ref{sec:cyclicKR}. In other words, the computational grid is too coarse in some regions to capture all qualitative features of the intermediate functions $\rho^{(n)}$, and as a result the transformations furnished by the method display visually apparent artifacts. 

By comparison, suppose for the sake of argument that we already knew
the right-hand side $\tilde{\rho}(\by)\propto\frac{1}{\rho(S(\by))}$
of (\ref{eq:invmae}). In fact $\tilde{\rho}(\by)$ is significantly
smoother than $\rho(\bx)$ because the composition of $\rho$ with
$S$ precisely smooths out the nonuniformity in $\rho$.

Thus if we apply the cyclic Knothe-Rosenblatt flow to the computation
of $S$ solving 
\begin{equation}
\det DS(\by)\propto\tilde{\rho}(\by),\label{eq:invmae2}
\end{equation}
 it is reasonable to use a separable grid with uniform resolution
to fit each successive right-hand side $\rho^{(n)}$ in the flow.

The only catch is that $\tilde{\rho}$ is not known \emph{a priori},
so it must be determined self-consistently. Therefore we propose the
following procedure, alternating the following steps until convergence
is reached:
\begin{enumerate}
\item Define $\tilde{\rho}(\by)\propto\frac{1}{\rho(S(\by))}$, fitting
with a sum of separable functions by evaluation of $S$ on a suitable
grid.
\item Construct $S$ solving (\ref{eq:invmae}) by cyclic Knothe-Rosenblatt
transport.
\end{enumerate}
In practice the convergence is rapid and requires less than 20 iterations.
We comment that the same self-consistent loop was used in the work of Lindsey and Rubinstein~\cite{LindseyRubinstein2017}
to solve general Monge-Amp\`ere equations of the form $\det DT(\bx)=f(\bx)/g(T(x))$,
by likewise reducing to the case where the right-hand side is independent
of $T$.

In fact, the cyclic Knothe-Rosenblatt flow of Section \ref{sec:cyclicKR},
in which the intermediate right-hand sides $\rho^{(n)}$ are fit by interpolating
on the uniform computational grid (via Chebyshev or Fourier interpolation),
can be viewed as a \emph{collocation} method for the Monge-Amp\`ere
equation (\ref{eq:mae}) in the sense that the solution $T$ solves
the PDE (\ref{eq:mae}) \emph{exactly} at the computational grid points.
Likewise, if the self-consistent procedure for computing $S=T^{-1}$
is solved to convergence, this method can be viewed as a collocation
method for (\ref{eq:invmae}), in the sense that $\det DS(\by)\propto1/\rho(S(\by))$
holds \emph{exactly} at the computational grid points in the $\by$
domain. Since $\rho(S(\by))$ has more uniform smoothness, this approach
is preferable as a collocation method.

In turn $T$ can be recovered if desired from the inversion procedure
outlined at the end of Section \ref{sec:cyclicKR}, if desired. However,
for downstream purposes all that we shall require will be the evaluation
on our uniform grid $\{\by_{n}\}$ of $S$, i.e., $S(\by_{n})=T^{-1}(\by_{n})=\bx_{n}$,
as well as of the full Jacobian matrix $DS(\by_{n})=\left[DT(\bx_{n})\right]^{-1}$,
which we obtain by analytical differentiation of the layers of Knothe-Rosenblatt as computed in Section~\ref{sec:separableKR}.

\begin{figure}
\noindent\begin{centering}\includegraphics[width=0.5\textwidth, trim={0 50 0 50},clip]{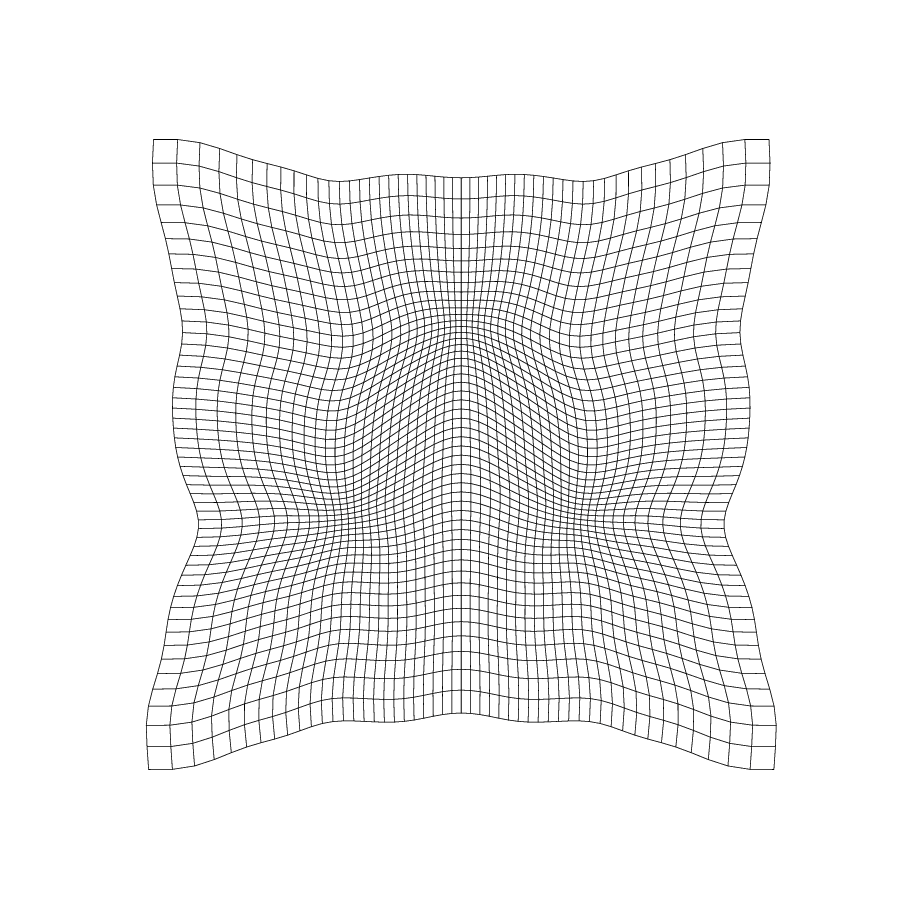}\includegraphics[width=0.5\textwidth, trim={0 50 0 50},clip]{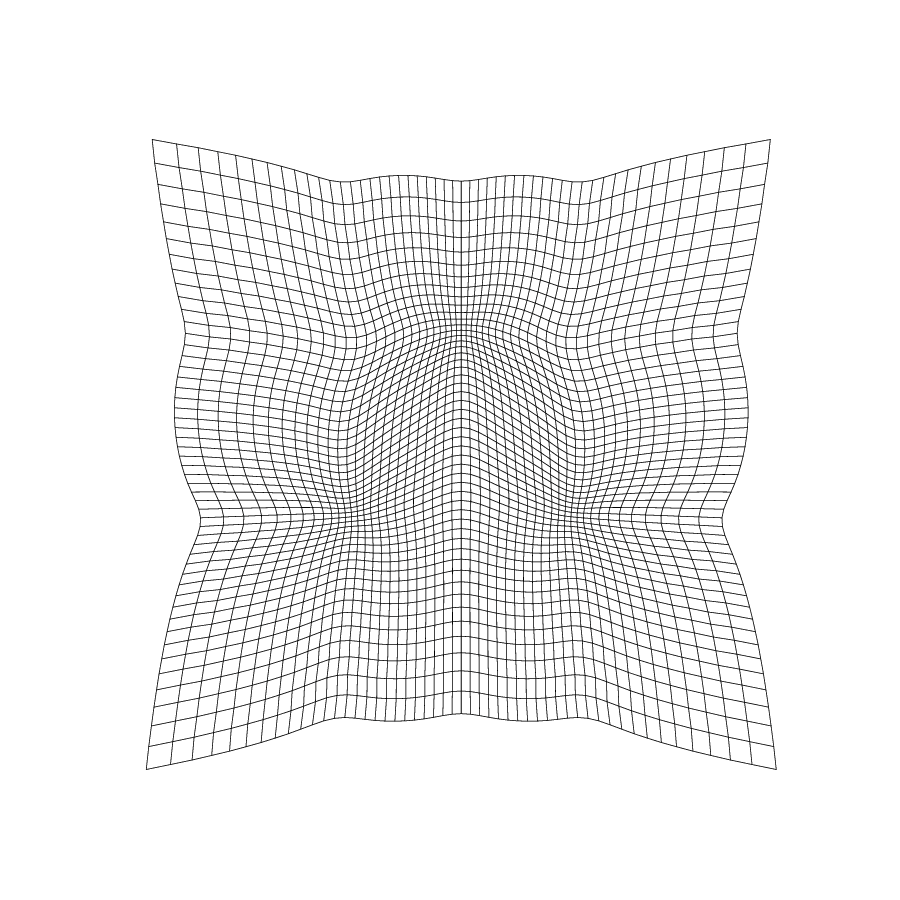}
\end{centering}
\caption{Image of the subset $[-4,4]^2 \times \{0\}$ of the computational domain $[-5,5]^3$ under $S = T^{-1}$, where $S$ is computed via the self-consistent procedure outlined in Section~\ref{sec:finalKR} and $\rho$ is the same as in Figure~\ref{fig:alias} Left: result using 21 computational grid points per dimension. Right: result using 65 computational grid points per dimension. Observe that, in contrast with Figure~\ref{fig:alias}, there is no noticeable aliasing error for the result with the coarser grid, which is almost converged to the exact limit.}
\label{fig:noalias}
\end{figure}

\section{Pseudospectral method for integral evaluation \label{sec:pseudo}}

Before describing how the integrals are evaluated, we provide a brief outline of the Hartree-Fock (HF) 
algorithm used in this work, in order to clarify our exact computational requirements. We refer
the reader to several excellent textbooks\cite{szabo1996modern,martin2020electronic,lin2019mathematical}.

\subsection{Review of SCF for HF equations} \label{sec:reviewHF}
The HF equations are solved using the self-consistent field (SCF) iteration. Within each 
iteration of the most basic approach, we perform the following two steps, which are then repeated over many iterations until convergence of the occupied molecular orbitals $f_p$ for $p=1,\ldots, N_e /2$ is achieved.
\begin{enumerate}
\item Given current guesses for the occupied molecular orbitals $f_p$ for $p=1,\ldots, N_e /2$, we construct a subroutine that can compute the action of the Fock operator on an arbitrary smooth function $f(\bx)$. The Fock operator is equal to the sum of kinetic ($\hat{T}$), nuclear
($\hat{N})$, Coulomb ($\hat{J}$), and exchange ($\hat{K}$) operators, whose actions are defined by:
\begin{align}
\hat{T}f (\bx) & =-\frac{1}{2} \Delta f (\bx)\nonumber \\
\hat{N}f (\bx) & =\sum_{I}V_{I}(\bx)f (\bx)\nonumber \\
\hat{J}f (\bx) & = \left[ \int \frac{1}{|\bx-\bx'|} \,  \rho(\bx') \,d\bx' \right]  f (\bx)  \nonumber \\
\hat{K}f (\bx) & =-\sum_{p=1}^{ N_{e} / 2}  f_{p}(\bx ) \int  \frac{1}{|\bx-\bx'|}f_{p}(\bx')f (\bx')\,d\bx'. \label{eq:4ops}
\end{align}
Here $V_{I}(\bx)$ is once again the nuclear potential due to the $I$-th nucleus, $N_e$ is the number of 
electrons, the index $p$ runs over occupied orbitals (each of which corresponds to two spin-orbitals), and $\rho(\bx)=2\sum_{p=1}^{N_{e}/2}\left|f_{p}(\bx)\right|^{2}$ 
is the electron density.  We will represent the function $f$ as well as the molecular orbitals $f_p$ via their evaluations $f (\bx_i)$ and  $f_p (\bx_i)$ on the 
deformed grid points, which can be related to coefficients in the deformed interpolating basis $\psi_i$ via \eqref{eq:psiinterp} or in the deformed orthonormal basis $\eta_i$ via \eqref{eq:etaInterp}. By `computing the action of the Fock operator,' we mean that given these representations, we can construct (within our pseudospectral approximation framework) the left-hand sides of \eqref{eq:4ops} evaluated at the same points $\bx = \bx_i$.
\item Given a subroutine computing the action of the Fock operator, we use the Davidson method\citep{davidson1975iterative} to obtain
the $N_e /2$ lowest eigenpairs of the Fock
operator. The eigenvectors define the next guesses $f_p$, $p=1,\ldots ,N_e / 2$, for the occupied molecular
orbitals.
\end{enumerate}

The basic 2-step
algorithm outlined above often fails to converge, and we use Pulay's DIIS\citep{pulay1980diis,pulay1982diis} 
to stabilize and accelerate the convergence. The computational bottleneck in these steps
is the application of the exchange operator. To minimize the number of exchange operator actions  
required, we use the adaptively compressed exchange (ACE) approach due
to Lin\citep{lin2016adaptively,hu2017adaptively}, which partitions the algorithm into two nested
loops. In the inner loop, the ACE exchange is kept fixed, and it is
only updated in the outer loop once the inner loops converges. Typically one needs to perform $O(10)$ iterations of the inner loop between for each outer loop iteration. In each iteration of the outer loop, we must solve $N_e^{2}$ Poisson
equations to construct the ACE operator. The overall cost of the
calculation will scale as $\tilde{O}(M N_e^2)$, where the tilde indicates the omission of log factors and $M$ is the number of basis functions.

In
this publication we will not focus on improving the convergence rate
of the SCF cycles or the Davidson solver. Instead we will only describe
how the various operator actions in (\ref{eq:4ops}) can be performed.

\subsection{General operator matrix elements} \label{sec:matrixelements}

As we will see below, the operator of central importance to us is the kinetic energy operator, because it is used not only to evaluate the kinetic energy but also to solve the Poisson equation for evaluating the Coulomb, exchange, and nuclear terms. It will also become apparent in Section~\ref{subsec:Kinetic-operator} that the matrix describing the action of the kinetic energy on the deformed sinc basis functions $\psi_i(\bx
)$  can be evaluated efficiently using FFTs.  However, because the $\psi_i(\bx)$ functions are not orthogonal, care is needed while using this matrix as we describe below.

For a general operator $\hat{O}$ (e.g., $-\Delta$ for the kinetic energy or $\Delta^{-1}$ for the solution operator of the Poisson equation), consider the matrix elements $O_{ij} = \langle \eta_i, \hat{O} \eta_j \rangle$ in our orthonormal basis. Here $\langle  \, \cdot \, , \, \cdot \, \rangle$ denotes the ordinary $L^2$ inner product over the computational domain. The matrix elements satisfy the defining property that for a basis expansion $f(\bx) = \sum_j f_j \, \eta_j (\bx)$ of some function $f$, we can expand $(\hat{O} f) (\bx)$ in the basis as $(\hat{O}f) (\bx) = \sum_i \left( \sum_j O_{ij} f_j \right) \eta_i (\bx)$. Equivalently, it holds that 
\begin{equation}
    \langle \eta_i, \hat{O} f \rangle = \sum_{j} O_{ij} f_j
    \label{eq:definingprop}
\end{equation}
We claim that the defining property \eqref{eq:definingprop} still holds for the alternative matrix elements 
\begin{equation}
    O_{ij} = \sqrt{J_i} \, \langle \psi_i, \hat{O} \, \psi_j \rangle \sqrt{J_j},
    \label{eq:altmatrix}
\end{equation}
provided that $f$ is suitably smooth relative to our deformed coordinates, as if we were effectively identifying $\eta_i (\bx) = \psi_i (\bx) \,  \sqrt{J(\bx)} $ with $\psi_i (\bx)\, \sqrt{J_i}$. The claim is justified in Appendix~\ref{sec:pseudocalc}.

Thus, it is easy to show that we can compute the action of $\hat{O}$ on a smooth function $f$ by simply performing matrix-vector multiplications by the matrix $\langle \psi_i, \hat{O} \, \psi_j \rangle$. To see this, recall that the coefficients $f_i$ in the $\{ \eta_i \}$ basis correspond to deformed grid point evaluations $f(\bx_i)$ via $f_i = w f(\bx_i) / \sqrt{J_i}$, and likewise $\langle \eta_i, \mathcal{O}f \rangle = w \, (\hat{O}f) (\bx_i) / \sqrt{J_i}$. 
Therefore we can alternatively express the action of $\hat{O}$ in terms of evaluations on the deformed grid points via the identity 
\begin{equation}
    (\hat{O} f) (\bx_i) = {J_i} \sum_j \langle \psi_i, \hat{O} \, \psi_j \rangle \, f( \bx_j )
    \label{eq:Ofgrid}
\end{equation}
from which the matrix elements $\langle \eta_i, \mathcal{O}f \rangle$ can be readily obtained from $(\hat{O} f) (\bx_i)$.

\subsection{Kinetic matrix \label{subsec:Kinetic-operator}}

Following the discussion in the preceding Section~\ref{sec:matrixelements}, we want to show how to multiply by the Laplacian matrix elements $\Lap_{ij} := - \langle \psi_i, \Delta \, \psi_j \rangle$. In Appendix~\ref{sec:fftkinetic}, we explain how this operation can be performed with pseudospectral accuracy, using 4 FFTs and 4 inverse FFTs.

The matrix-vector multiplication by $\Lap$ not only allows us to perform the action of the kinetic operator $\hat{T}$, but also will serve as a key computational primitive that we leverage in the calculation of both nuclear and two-electron integrals. Both of these types of integral can be understood in terms of the Poisson equation, which we shall discuss first in an abstract context.

\subsection{Solving the Poisson equation\label{subsec:Solving-Poisson-equation}}

In this section we explain how to use the Laplacian matrix $\Lap$ to solve a general Poisson equation
\begin{equation}
\label{eq:Poisson}
    -\Delta u (\bx) = v (\bx)
\end{equation}
within the pseudospectral approximation, for general right-hand side $v(\bx)$. The method is motivated by stipulating that the Poisson equation~\eqref{eq:Poisson} holds on the deformed grid points, i.e.,
\begin{equation}
\label{eq:Poissongrid}
    -\Delta u (\bx_i) = v (\bx_i)
\end{equation}
for all $i$.

To solve \eqref{eq:Poissongrid}, take $\hat{O} = -\Delta$ in \eqref{eq:Ofgrid} to obtain equations 
\begin{equation}
\label{eq:poissonpointwise}
     {J_i} \sum_j \Lap_{ij}\, u(\bx_j) = v(\bx_i),
\end{equation}
which are equivalent to \eqref{eq:Poissongrid} up to the pseudospectral approximation. 
Therefore we can solve the linear system 
\begin{equation}
\label{eq:linsys}
    \Lap \tilde{u} = b,
\end{equation}
where the right-hand side is defined by $b_i = v(\bx_i)/J_i$, and the solution vector $\tilde{u}_j = (u(\bx_j) )$ is the vector of evaluations of the solution $u(\bx)$ of the Poisson equation~\eqref{eq:Poisson} on the deformed grid points. In turn $u(\bx)$ can be recovered in either the $\{ \psi_i \}$ or $\{ \eta_i \}$ basis via interpolation, cf.~\eqref{eq:psiinterp} and \eqref{eq:etaInterp}.

Since $\Lap$ is positive definite by construction, we can solve the linear system \eqref{eq:linsys} using the conjugate gradient (CG) method, relying on the subroutine discussed in Section~\ref{subsec:Kinetic-operator} to perform matrix-vector multiplications by $\Lap$ without every forming the matrix fully. For our preconditioner we use the exact inverse $\tilde{\Lap}^{-1}$ of the `undeformed' Laplacian matrix $\tilde{\Lap}_{ij} = \langle \phi_i, -\Delta \phi_j \rangle$, which is diagonalized by the discrete Fourier transform. Thus the predconditioner $\tilde{\Lap}^{-1}$ can be applied to a vector using a single FFT and a single inverse FFT. In our experiments we require on roughly 20 to 30 CG iterations
to solve \eqref{eq:linsys}. Since each iteration requires 5 forward and inverse FFTs, the solve is about 100 times more expensive than the analogous solve for a uniformly spaced
sinc basis.

\subsection{Coulomb and exchange operators}
\label{sec:coul_exch}
The action of the Coulomb and exchange operators $\hat{J}$ and $\hat{K}$ defined in in \eqref{eq:4ops} can be computed by leveraging our subroutine for solving the Poisson equation outline in the preceding Section~\ref{subsec:Solving-Poisson-equation}. Indeed, the action of the integral Coulomb kernel $\frac{1}{\vert \bx - \bx' \vert }$ appearing in the definitions of $\hat{J}$ and $\hat{K}$ can be viewed (up to a proportionality constant) as a Poisson solve. In fact, the formulas of \eqref{eq:4ops} involving the Coulomb kernel are valid only for open boundary conditions, but more generally the understanding of these operations as Poisson solves holds even when periodic boundary conditions are considered, provided that the Laplacian is considered with appropriate boundary conditions.

To apply $\hat{J}$, we must form the electron density values $\rho (\bx_j)$ on our deformed grid and then performs the Poisson solve 
\begin{equation*}
    \Delta u (\bx) = \rho(\bx)
\end{equation*}
following the method of Section~\ref{subsec:Solving-Poisson-equation}, which recovers the values of the solution $u(\bx_i)$ on the deformed grid points. Then $\hat{J} f (\bx_i)$ can be constructed from $u(\bx_i)$ and $f(\bx_i)$ by pointwise multiplication on the deformed grid.

To form $\hat{K} f (\bx_i)$, the logic is much the same, except that we must perform $N_e / 2$ Poisson solves:
\begin{equation*}
    \Delta u_p (\bx) = f_p (\bx) f(\bx) 
\end{equation*}
to furnish $u_p (\bx_i)$ for $p=1,\ldots , N_e/2$.

\subsection{Pseudopotential-based nuclear operator}
\label{sec:ppnuc}
The action of the nuclear operator $\hat{N}$ in \eqref{eq:4ops} is quite simple to perform under the assumption that the nuclear potentials $V_I$ are smooth relative to our deformed coordinates. Note that this is not exactly the case for the bare nuclear potential $V_I (\bx) = -Z_I / \vert \bx - R_I \vert$, due to the singularity at the atom center. By considering more strongly deformed coordinates near the origin, we improve the quantitative smoothness near the center but cannot remove the singularity. Thus we are motivated to introduce an all-electron pseudopotential (PP) following the recent work of Gygi\citep{Gygi2023}. For the purposes of this section, we view $V_I$ as smooth, having been constructed as a PP following the discussion in Appendix~\ref{sec:PP}, and in this construction we only need to assume that $V_I (\bx_i)$ can be evaluated on our deformed grid. The details of an alternate approach using the bare nuclear potential will be provided in the following Section~\ref{sec:bare_nuc}.

Under the PP assumptions outlined above, the action of the nuclear operator $\hat{N}$ defined in \eqref{eq:4ops} can be performed simply as 
\begin{equation*}
\hat{N} f (\bx _i ) = \sum_I V_I (\bx_i) f(\bx_i).
\end{equation*}

\subsection{Bare nuclear operator}
\label{sec:bare_nuc}

Consider a general diagonal operator $\hat{V}$ which operates as $\hat{V} f (\bx) = V(\bx) f(\bx)$. Compared to the discussion of Section~\ref{sec:matrixelements}, we will consider an alternative defining property for the matrix elements $V_{ij}$. Namely, given expansions $f(\bx) = \sum_i f_i \, \eta_i (\bx)$ and $g(\bx) = \sum_i g_i \, \eta_i (\bx)$ for suitably smoooth functions $f$ and $g$ in the orthonormal $\{ \eta _i \}$ basis, we insist that 
\begin{equation}
    \langle f, \hat{V} g \rangle = \int f(\bx) g(\bx) V(\bx) \, d\bx = \sum_{ij} f_{i}  g_{j} V_{ij}.
\end{equation}
By the same manipulations as in Section~\ref{sec:diagonaldef} (cf. \eqref{eq:boldu}, \eqref{eq:boldv}, and \eqref{eq:uvtilde}, this property is satisfied for either the choice $V_{ij} = \mathbf{v}_i \, \delta_{ij}$ or $V_{ij} = \tilde{\mathbf{v}}_i \, \delta_{ij}$, where $\bold{v}_i = \frac{\sqrt{J_i}}{w} \int \eta_i (\bx) \,   V(\bx) \, d \bx $ or $\tilde{\bold{v}}_i = \frac{J_i}{w} \int \psi_i (\bx) \,  V(\bx) \, d \bx $. We choose the latter as the more convenient, leading to formulation
\begin{equation}
\label{eq:barenucaction}
    \hat{N} f (\bx_i) = \sum_I V_{I,i} \, f(\bx_i),
\end{equation}
for the action of the nuclear operator \eqref{eq:4ops}, where 
\begin{equation}
\label{eq:VIi}
    V_{I,i} := \frac{J_i}{w} \int \psi_i (\bx) \,  V_I (\bx) \, d\bx
\end{equation}
can be evaluated even in the case of the bare nuclear potential $V_I = -Z_I / \vert \bx - R_I \vert$, which has an integrable singularity. Note that the formulation \eqref{eq:barenucaction} for the nuclear action defined in \eqref{eq:4ops} holds, via our defining property, in the sense of being accurate up to projection onto the space of suitably smooth functions.

In Appendix~\ref{sec:bare_nuc_comp}, we demonstrate how the calculation of $\sum_I V_{I,i}$ as in \eqref{eq:VIi} can be reduced to Poisson solves and in turn a single linear solve involving the matrix $\Lap$.

\section{Results}\label{sec:numerics}

Here we present the results for calculations on He, He$_{2}$, Be,
Be$_{2}$, and CH$_{4}$ systems. All calculations are performed on a cubic unit cell of size $10$ Bohr with periodic boundary conditions. 

For each system we compute a single transformation $T$ by constructing the self-consistent inverse cyclic Knothe-Rosenblatt flow following the discussion of Section~\ref{sec:finalKR}. The choice of $\rho$ that determines the transformation will be outlined below for each system. Then the same transformation $T$ is used with various choices $M$ of deformed sinc basis set size. For He, He$_{2}$, Be, and Be$_{2}$, we compute $T$ using a $20 \times 20 \times 20 $ computational grid. We use a $30 \times 30 \times 30$ computational grid to compute the transformation for CH$_4$. 

\subsection{He and He$_{2}$}

\begin{figure}
\noindent \begin{centering}
\includegraphics[width=0.49\textwidth]{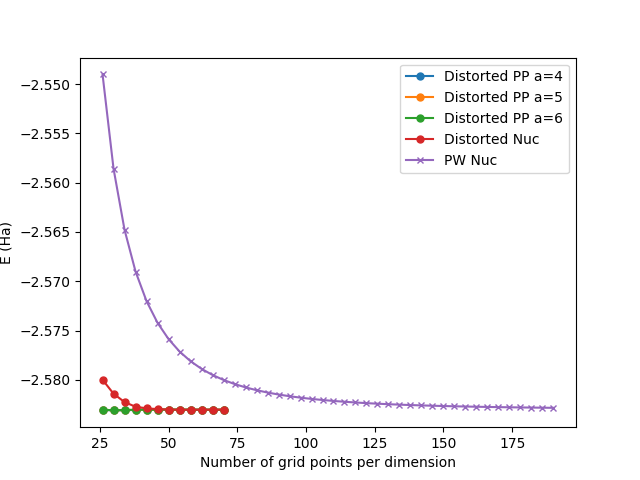}\hfill{}%
\includegraphics[width=0.49\textwidth]{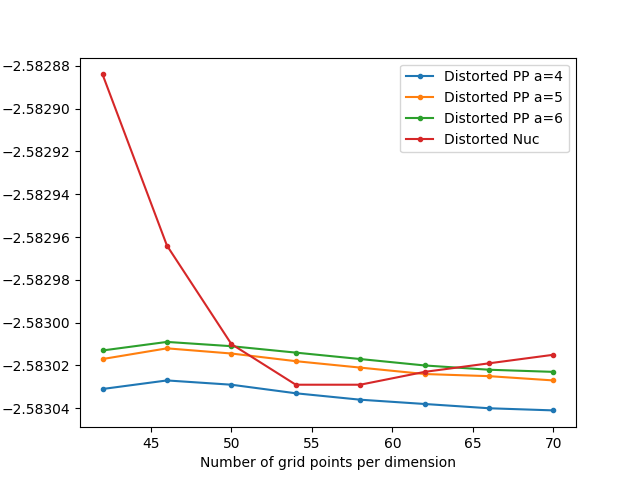}\\
\begin{tabular}{cc}
\hline 
Potential & Energy\tabularnewline
\hline 
PP, $a=4$ & -2.58304\emph{2}\tabularnewline
PP, $a=5$ & -2.58302\emph{7}\tabularnewline
PP, $a=6$ & -2.58302\emph{3}\tabularnewline
Bare Nuclear & -2.5830\emph{15}\tabularnewline
\hline 
\end{tabular}
\par\end{centering}
\caption{The left figure shows the convergence of the energy of He atom as the
number of basis functions is increased. The four curves with the
label ``distorted'' show the results obtained with distorted sinc
functions using both the PP (with parameter values $a=4,5,6$) and the bare nuclear potential. The difference between the energies is not visible on the scale
of this graph. The right graph is a zoomed-in version of the left graph where the convergence of the various curves is more clearly visible. For additional clarity the numerical values of the converged energies obtained with $M = 70^3$ basis functions are given in the table on the right. The italicized number corresponds
to the digit which is uncertain in our calculations due to the finiteness of the basis set.
The purple curve at left plots the convergence of energy when plane waves (PWs) are used together with the
bare nuclear potential. Points on the purple curve are sufficiently far away from convergence that they would fall outside of the range of values displayed on the right curve. \label{fig:tabconvhe}}
\end{figure}

For He and He$_{2}$ we prescribe $J(\bx)  \propto \rho(\bx)$ 
where 
\begin{equation}
\rho(\bx)=\sum_{I}\frac{\mathrm{erf}(Z_{I}|\bx-R_{I}|/a_{I})-\mathrm{erf}(Z_{I}|\bx-R_{I}|/b_{I})}{|\bx-R_{I}|}+c\label{eq:deform}.
\end{equation}
Here $R_{I}$ and $Z_{I}$ are again the position and charge of the $I$-th
nucleus, and $a_I=0.1$, $b_I=4$, and $c=0.01$ are parameters that determine
the spatial extent of the deformation. 
This deformation ensures that the density
of grid points increases approximately as $1/r$ between the distances
of $b/Z_{I}$ Bohr and $a/Z_{I}$ Bohr from the center of the nucleus. Outside of this region, the density is roughly a non-zero constant.
The total number of distorted sinc functions $M$ is progressively increased until convergence
is achieved. The Monge-Amp\`{e}re equation \eqref{eq:mae} is solved by the method of Section~\ref{sec:finalKR} using $N=15$ steps and $20^{3}$ computational grid points in each of the $N=15$ intermediate fitting steps.

Figure \ref{fig:tabconvhe} concerns the He atom exclusively and is used to justify the accuracy of our practical choice of all-electron pseudopotential (PP) for the remaining calculations. The numbers in the table in Figure \ref{fig:tabconvhe} show that the energies
of the pseudopotential (PP) calculations converge to that of the bare nuclear calculations as the PP parameter $a$ is increased. In all the calculations shown in the table we
have used a large number ($M = 70^{3}$) of basis functions, and our estimated uncertainty
due to basis set incompleteness with PP are on the order of a few $\mu$E$_{h}$.
The table shows that when $a=4$ is used in the PP we can expect to
get an energy which is within a few tens of $\mu$E$_{h}$ of the
converged result compared to the bare nuclear potential. The left
graph in Figure (\ref{fig:tabconvhe}) shows that the convergence
of the energy with distorted grids is vastly improved compared
the convergence with undistorted grids. One would expect that the difference
in performance would increase for larger atoms, as all-electron calculations with plane waves for heavier atoms quickly become impractical without the use of large supercomputers. The right graphs shows the zoomed in version of the left graph where we have not plotted the energies with the undistorted grids. This graphs shows that the converge of energy is much more smoother when PP is used and the bare nuclear energies are still not fully converged to $\mu$E$_h$ even with $M = 70^{3}$ number of basis functions.

\begin{figure}
\begin{centering}
\includegraphics[width=0.49\textwidth]{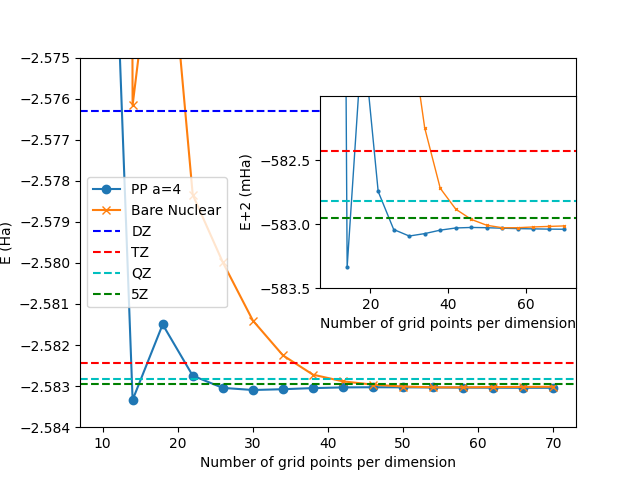}\hfill{}\includegraphics[width=0.49\textwidth]{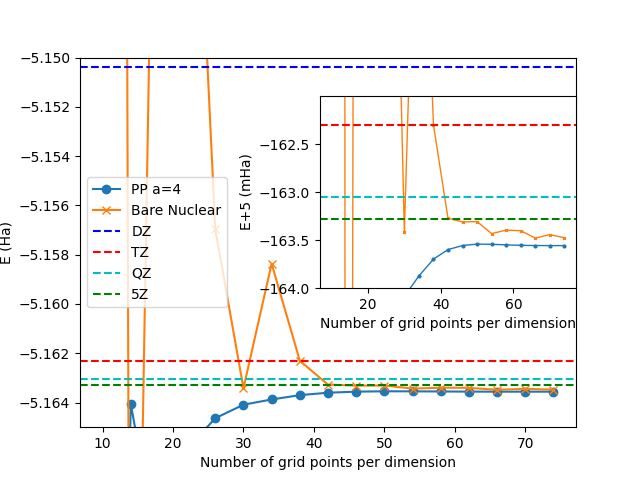}\caption{The figure on the left (resp., right) shows the convergence of the energy with respect to the distorted basis set size for the He atom (resp., He$_{2}$ molecule), with comparisons to GTO energies plotted as horizontal lines. \label{fig:heconv}}
\par\end{centering}
\end{figure}

Figure (\ref{fig:heconv}) shows that we can converge the energy of He atom and
He dimer (with a bond length of 2 \AA) to accuracy surpassing that of the 5Z basis by using $30^{3}$ and $40^{3}$ grid
points, respectively, and the PP with $a=4$. Interestingly, the He dimer requires less than double the number of basis functions compared to the He atom, likely because
the density contributions in $\rho(\bx$), cf. \eqref{eq:deform}, of the two atoms overlap. It is
also worth noting that the convergence of the energy with respect to the number of grid points is significantly less smooth for the bare nuclear
potential than for the PP.

\subsection{Be and Be$_{2}$}

\begin{figure}
\centering{}\includegraphics[width=0.49\textwidth]{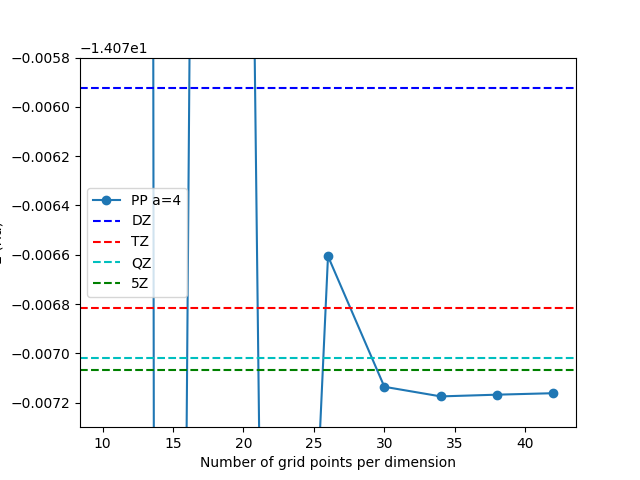}\hfill{}\includegraphics[width=0.49\textwidth]{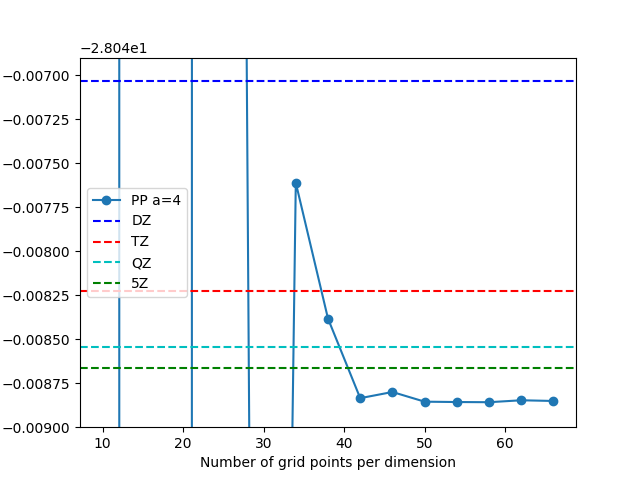}\caption{The figure on the left (resp., right) shows the convergence of the energy with respect to the distorted basis set size for the Be atom (resp., Be$_{2}$ molecule), with comparisons to GTO energies plotted as horizontal lines.  \label{fig:beconv}}
\end{figure}

Next we use the procedure outlined above to evaluate the energy of the
beryllium atom and beryllium dimer (with a bond length of 2.44 \AA). For these
systems we have used the same deformation as (\ref{eq:deform}) with
$a_I=0.1$, $b_I=8$, and $c=0.03$. Figure (\ref{fig:beconv}) shows that the energy surpasses the accuracy of the 5Z basis set with grid
sizes of $30^{3}$ and $42^{3}$, respectively, for the atom and the
dimer. These numbers are not significantly different than the ones
we found for He atom, suggesting that with an appropriately
chosen deformation, heavy nuclei can be treated without significant
overhead. Further tests will be needed to confirm this claim for a range of heavier atoms. We point out that, once again, the number of grid points needed
for the dimer is only slighly higher than the number needed for the atom because the density contributions of the two nuclei in the specification of $\rho(\bx)$ have significant overlap.

\subsection{CH$_{4}$}

\begin{figure}
\centering{}\includegraphics[width=0.49\textwidth]{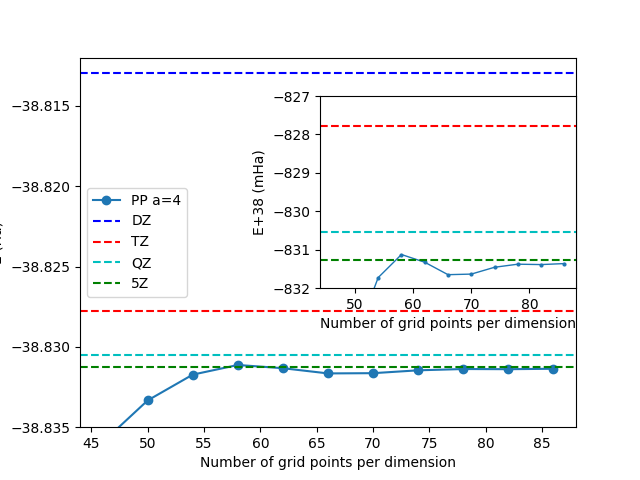}\caption{The figure shows the convergence of the energy with respect to the distorted basis set size for the methane molecule.  \label{fig:ch4conv}}
\end{figure}

For our final test we perform a calculation on methane, 
which does not have a linear geometry, suggesting that simpler approaches based on 1-D deformations will not be highly efficient. 
Again we use the deformation (\ref{eq:deform}), choosing $a_I=0.1, \, b_I=18$ for the carbon atom and $a_I=0.1, \, b_I=1.5$ for the hydrogen atoms, as well as $c=0.01$.
These parameters are chosen to ensure that a higher density of basis functions 
surrounds the carbon atom, which is bonded to all the hydrogen atoms
surrounding it. Results are plotted in Figure~\ref{fig:ch4conv}. For this system we observe that for $M = 50^{3}$ grid points, the energy is converged to within a few tenths of mE$_{h}$.
We can further converge the results within a few tens
of $\mu$E$_{h}$ by taking $M = 80^{3}$ grid points. 

\section{Discussion and conclusion}

In this article we have shown how the two major challenges associated with
the use of the distorted sinc basis set can be overcome. These challenges are, namely: (1) computing the
transformation maps that yield a prescribed density of basis functions and (2) calculating and operating with integrals for the Hamiltonian. We have demonstrated that these techniques can be used to
evaluate mean-field energies at a cost that is log-linear in the number of basis functions.

Several questions motivate further investigation in this direction.
First, in this work we have designed the deformation by prescribing a density of basis functions with an analytical functional form. A more systematic approach with less hand-tuning would be desirable. Such an approach might be achieved 
by minimizing the variance of energy for a fixed number of grid points. (One cannot directly minimize the energy because it is not variational
due to the use of diagonal approximation.) Another approach might be to minimize
the error in the one- and two-electron integrals for GTO basis functions incurred by representing the GTOs in the interpolating basis of deformed sinc functions.

Second, we expect to be able to get away with a smaller number
of basis functions if we can combine the deformed sinc basis set with sharp GTO
basis functions to describe the nuclear cusp. This perspective has been applied fruitfully in several works~\cite{White2023,GaviniEnriched1,dft-fe-1.0}.
It is an open question as to whether such a mixed basis set can be used while keeping the cost of the calculation linear in the number
of basis functions as we have done here.

Third, currently the most significant cost of the calculation is the solution of the Poisson equation. The use of the preconditioner as described in Section \ref{subsec:Solving-Poisson-equation} significantly speeds up the calculations, but we still require on the order of a few 10's of CG iterations per solve. We plan to investigate more sophisticated proconditioners in future work. 

Finally, we plan to investigate applications of our basis set in frameworks for correlated calculations, including tensor networks methods, auxiliary field quantum Monte Carlo, variational Monte Carlo, and quantum algorithms. 

\section{Acknowledgements}

The authors would like to thank Francois Gygi and Steve White for
helpful discussions. S.S. was supported through the National Science
Foundation grant CHE-2145209. M.L. was supported by the Applied Mathematics Program of the US Department of Energy (DOE) Office of Advanced Scientific Computing Research under contract number DE-AC02-05CH11231.

\bibliographystyle{unsrt}

\newpage{}

\appendix

\section{Periodic sinc basis}

\subsection{Definitions}
\label{sec:sincdefs}

Let $\bL=(L_{1},L_{2},L_{3})$ be the unit cell side lengths, and
let $\mathcal{V}=L_{1}L_{2}L_{3}$ denote the volume of the unit cell.
Let $\bm=(m_{1},m_{2},m_{3})$ denote the numbers of grid points per
dimension. We assume that these numbers are odd in order to construct
the periodic sinc basis. To index the Fourier basis, it is useful
then to define the integers $\ell_{a}=(m_{a}-1)/2$ for $a=1,2,3$.
We also let $M=m_{1}m_{2}m_{3}$ denote the total number of basis
functions and $\Delta\mathcal{V}=\mathcal{V}/M$ the discrete volume
element.

For the purposes of this discussion, let us identify the indices $i,j$
for the grid points / basis functions with the multi-indices $\bi,\bj\in\prod_{a=1}^{3}\{0,\ldots,m_{a}-1\}$.

Then the orthonormal periodic sinc basis functions can be expressed
\begin{equation}
\phi_{\bj}(\by)=\frac{1}{\sqrt{M}}\sum_{\bk}e^{-2\pi i\bk\cdot(\bj/\bm)}e_{\bk}(\by),\label{eq:sincfourier}
\end{equation}
 where vector division is performed entrywise and $e_{\bk}$ are the
orthonormal Fourier modes on our unit cell, defined by 
\[
e_{\bk}(\by)=\frac{1}{\sqrt{\mathcal{V}}}\,e^{2\pi i\bk\cdot(\by/\bL)},\quad\bk\in\prod_{a=1}^{3}\{-\ell_{a},\ldots,\ell_{a}\}.
\]
 Note the indexing convention for $\bk$.

\subsection{Kinetic matrix multiplication} \label{sec:fftkinetic}

Here we retain the multi-indexing convention of Appendix~\ref{sec:sincdefs}, to which we refer the reader for the detailed specification of the sinc basis set $\phi_{\bj}$.

Now we concern ourselves with demonstrating how to perform matrix-vector multiplications by the kinetic matrix $\Lap_{\bj \bj'} = -\int \psi_{\bj} (\bx) \Delta \psi_{\bj'} (\bx) \,dx$. Substituting $\psi_{\bj}:=\phi_{\bj}\circ T$, we compute: 
\begin{align}
\Lap_{\bj \bj'} & =-\int\psi_{\bj}(\bx)\Delta\psi_{\bj'}(\bx)\,d\bx\nonumber \\
 & =\int\nabla\psi_{\bj}(\bx)\cdot\nabla\psi_{\bj'}(\bx)\,d\bx\nonumber \\
 & =\int\left[\nabla\phi_{\bj}\circ T\right](\bx)\,\left[DT(\bx)\,DT(\bx)^{\top}\right]\,\left[\nabla\phi_{\bj'}\circ T\right](\bx)\,d\bx\nonumber \\
 & =\int\nabla\phi_{\bj}(\by)\,\left[\frac{K(\by)K(\by)^{\top}}{J(\by)}\right]\,\nabla\phi_{\bj'}(\by)\,d\by\nonumber \\
 & =\int\nabla\phi_{\bj}(\by)\,\Sigma(\by)\,\nabla\phi_{\bj'}(\by)\,d\by,\label{eq:tjjexpand}
\end{align}

where we have defined $K(\by) := [DT^{-1} (\by)]^{-1}$ and in turn $\Sigma(\by):=\frac{K(\by)K(\by)^{\top}}{J(\by)}$.

Next substitute the Fourier expansions (\ref{eq:sincfourier}) for
$\phi_{\bj}=\overline{\phi_{\bj}}$ and $\phi_{\bj'}$ into (\ref{eq:tjjexpand}),
using the identity $\nabla e_{\bk}(\by)=2\pi i\frac{\bk}{\bL}e_{\bk}(\by)$,
to obtain: 
\begin{equation}
\Lap_{\bj\bj'}=\frac{(2\pi)^{2}}{M}\sum_{\bk}e^{2\pi i\bk\cdot(\bj/\bm)}\left[\frac{\bk}{\bL}\right]^{\top}\sum_{\bk'}\hat{\Sigma}_{\bk\bk'}\left[\frac{\bk'}{\bL}\right]e^{-2\pi i\bk'\cdot(\bj'/\bm)},\label{eq:tjjexpand2}
\end{equation}
 where 
\[
\hat{\Sigma}_{\bk\bk'}:=\int\overline{e_{\bk}(\by)}\Sigma(\by)e_{\bk'}(\by)\,d\by.
\]

Now we will rewrite the integral $d\by$ as a Riemann sum over the
real space grid points $\by_{\bj}=\left(\frac{j_{a}}{m_{a}}L_{a}\right)$:
\begin{equation}
\hat{\Sigma}_{\bk\bk'}\approx\mathcal{V}\,\sum_{\bi}e^{-2\pi i\bk\cdot(\bi/\bm)}\,\Sigma(y_{\bi})\,M^{-1}e^{2\pi i\bk'\cdot(\bi/\bm)}.\label{eq:Sigmahat}
\end{equation}
 Note that $\hat{\Sigma}_{\bk\bk'}$ is $(3\times3)$-matrix-valued.
We have split the volume element $\Delta\mathcal{V}=\mathcal{V}/M$
into pieces to more naturally suggest forward and inverse discrete
Fourier transforms.

Now we are ready to describe how to perform a matrix vector product
of $(\Lap_{\bj\bj'})$ by $(u_{\bj'})$. Observe first that from (\ref{eq:tjjexpand2}),
we have 
\[
\sum_{\bj'}\Lap_{\bj\bj'}v_{\bj'}=\frac{(2\pi)^{2}}{M}\sum_{\bk}e^{2\pi i\bk\cdot(\bj/\bm)}\left[\frac{\bk}{\bL}\right]^{\top}\sum_{\bk'}\hat{\Sigma}_{\bk\bk'}\left[\frac{\bk'}{\bL}\right]\hat{u}_{\bk'},
\]
 where (subject to suitable indexing) $\hat{u}$ can be obtained precisely
as an FFT of $u$. (We have summed over $\bj'$.)

Then we form the 3-vector-valued Fourier grid function 
\[
\left(\hat{v}_{\bk'}^{b}\right)=\left(\left[k'_{b}/L_{b}\right]\hat{u}_{\bk'}\right)
\]
 by pointwise operations over the Fourier grid.

In turn, to compute $\hat{w}_{\mathbf{k}}:=\sum_{\bk'}\hat{\Sigma}_{\bk\bk'}\hat{v}_{\bk'}$,
defined elementwise as 
\[
\hat{w}_{\bk}^{a}=\sum_{b,\bk'}\hat{\Sigma}_{\bk\bk'}^{ab}\hat{v}_{\bk'}^{b}
\]
 for $a=1,2,3$, we insert into (\ref{eq:Sigmahat}). Evidently this
operation can be achieved with one inverse FFT per coordinate $b=1,2,3$,
followed by a pointwise $(3\times3)\times(3\times1)$ matrix-vector
multiplication at each real space grid point, followed by one forward
FFT per coordinate $a=1,2,3$.

Finally, we can recover 
\[
\sum_{\bj'}\Lap_{\bj\bj'}v_{\bj'}=\frac{(2\pi)^{2}}{M}\sum_{\bk}e^{2\pi i\bk\cdot(\bj/\bm)}\left[(\bk/\bL)\cdot\hat{w}_{\bk'}\right]
\]
 with more pointwise operations on the Fourier grid, followed by one
final inverse FFT.

\section{Knothe-Rosenblatt computations}

\subsection{Periodic extension \label{sec:periodicboundary}}

We assume that $\rho$ extends smoothly and periodically from the
box $\mathcal{B}$ to $\R^{d}$, and we want to show that the Knothe-Rosenblatt
transport $T:\mathcal{B}\ra\mathcal{B}$ extends smoothly and periodically---in
the sense that the coordinate functions of $T-\mathrm{Id}$ are all
periodic---to a map $\R^{d}\ra\R^{d}$ that preserves each unit cell.

To see this, simply adopt the same definitions (\ref{eq:T1}), (\ref{eq:T2}),
(\ref{eq:T3}) for $T_{1},T_{2},T_{3}$, but extend their domains
to $\R^{d}$. Then evidently the map $T$ is a smooth map $\R^{d}\ra\R^{d}$.
It remains to demonstrate that the extension preserves the boundary
of all unit cells.

First it is useful to note that the functions $\rho_{1}(\,\cdot\,)$,
$\rho_{2}(\,\cdot\,\vert x_{1})$, etc. are respectively $L_{1}$-periodic,
$L_{2}$-periodic, etc. To see this first observe that the marginals
$\rho_{1}(x$

For $k_{1}\in\mathbb{Z}$, 
\begin{align*}
T_{1}(a_{1}+k_{1}L_{1}) & =a_{1}+L_{1}\int_{a_{1}}^{a_{1}+k_{1}L_{1}}\rho_{1}(y_{1})\,dy_{1}\\
 & =a_{1}+L_{1}\sum_{j_{1}=0}^{k_{1}-1}\int_{a_{1}+j_{1}L_{1}}^{a_{1}+(j_{1}+1)L_{1}}\rho_{1}(y_{1})\,dy_{1}\\
 & \overset{(\star)}{=}a_{1}+L_{1}\sum_{j_{1}=0}^{k_{1}-1}\int_{a_{1}}^{a_{1}+L_{1}}\rho_{1}(y_{1})\,dy_{1}\\
 & =a_{1}+k_{1}L_{1}\int_{a_{1}}^{a_{1}+L_{1}}\rho_{1}(y_{1})\,dy_{1}\\
 & =a_{1}+k_{1}L_{1},
\end{align*}
 where the $(\star)$ step follows from the fact that $\rho_{1}$
is $L_{1}$-periodic.

Similarly, for $k_{2}\in\mathbb{Z}$, 
\begin{align*}
T_{2}(x_{1},a_{2}+k_{2}L_{2}) & =a_{2}+L_{2}\int_{a_{2}}^{a_{2}+k_{2}L_{2}}\rho_{2}(y_{2}\vert x_{1})\,dy_{2}\\
 & =a_{2}+L_{2}\sum_{j_{2}=0}^{k_{2}-1}\int_{a_{2}+j_{2}L_{2}}^{a_{2}+(j_{2}+1)L_{2}}\rho_{2}(y_{2}\vert x_{1})\,dy_{2}\\
 & =a_{2}+L_{2}\sum_{j_{2}=0}^{k_{2}-1}\int_{a_{2}}^{a_{2}+L_{2}}\rho_{2}(y_{2}\vert x_{1})\,dy_{2}\\
 & =a_{2}+k_{2}L_{2}.
\end{align*}
 Note that we have used the fact that $\rho_{2}(\,\cdot\,\vert x_{1})$
is $L_{2}$-periodic.

Extrapolating the same argument to $T_{3}$, etc., establishes that
$T$ preserves the boundary of all unit cells.

\subsection{Explicit solution for a sum of separable functions \label{sec:separable}}

Suppose $\rho$ has the functional form 

\[
\rho(\bx)=\sum_{\alpha}c^{\alpha}g^{\alpha}(\bx),
\]
 where each term $g^{\alpha}(\bx)=\prod_{i=1}^{d}g_{i}^{\alpha}(x_{i})$
is separable. Let $G_{i}^{\alpha}(x_{i})=\int_{a_{i}}^{x_{i}}g_{i}^{\alpha}(y_{i})\,dy_{i}$
be the appropriate antiderivative, and assume that we can evaluate
$g_{i}^{\alpha},G_{i}^{\alpha}$ easily. We will show how to compute
the Knothe-Rosenblatt transport in terms of such evaluations.

Since $\rho$ must be a density on $\mathcal{B}$, we must have that
\[
1=\int_{\mathcal{B}}\rho\,d\bx=\sum_{\alpha}c^{\alpha}\prod_{i=1}^{d}G_{i}^{\alpha}(b_{i})=1.
\]
 Accordingly define the univariate masses 
\[
m_{i}^{\alpha}=G_{i}^{\alpha}(b_{i})=\int_{a_{i}}^{b_{i}}g_{i}^{\alpha}\,dx_{i}
\]
 and 
\[
m^{\alpha}:=\prod_{i=1}^{d}m_{i}^{\alpha}=\int_{\mathcal{B}}g^{\alpha}\,d\bx,
\]
 so we need 
\[
\sum_{\alpha}m^{\alpha}c^{\alpha}=1.
\]
 Therefore, given unnormalized $\tilde{c}^{\alpha}$, we can define
normalized weights
\[
c^{\alpha}=\frac{\tilde{c}^{\alpha}}{M},\quad\text{where }M:=\sum_{\alpha}m^{\alpha}\tilde{c}^{\alpha},
\]
 ensuring that $\rho$ is a valid probability density.

It is further useful to define 
\[
\omega_{i}^{\alpha}:=\prod_{j=i+1}^{d}m_{i}^{\alpha}.
\]

We want to evaluate the expressions above for $T_{k}$. To begin,
the marginal $\rho_{1}$ can be computed as 
\[
\rho_{1}(x)=\sum_{\alpha}c^{\alpha}\omega_{1}^{\alpha}g_{1}^{\alpha}(x).
\]
 Then compute: 
\begin{align*}
T_{1}(x_{1}) & =a_{1}+L_{1}\sum_{\alpha}c^{\alpha}\omega_{1}^{\alpha}G_{1}^{\alpha}(x_{1}).
\end{align*}

Next, simplify: 
\begin{align*}
T_{2}(x_{1},x_{2}) & =a_{2}+L_{2}\int_{a_{2}}^{x_{2}}\rho_{2}(y_{2}\vert x_{1})\,dy_{2}\\
 & =a_{2}+\frac{L_{2}}{\rho_{1}(x_{1})}\int_{a_{2}}^{x_{2}}\rho_{2}(x_{1},y_{2})\,dy_{2},
\end{align*}
 and compute 

\[
\rho_{2}(x_{1},x_{2})=\sum_{\alpha}c^{\alpha}\omega_{2}^{\alpha}g_{1}^{\alpha}(x_{1})g_{2}^{\alpha}(x_{2}).
\]
 Therefore 
\[
T_{2}(x_{1},x_{2})=a_{2}+\frac{L_{2}}{\rho_{1}(x_{1})}\left[\sum_{\alpha}c^{\alpha}\omega_{2}^{\alpha}g_{1}^{\alpha}(x_{1})G_{2}^{\alpha}(x_{2})\right],
\]
 or 
\[
T_{2}(x_{1},x_{2})=a_{2}+L_{2}\frac{\sum_{\alpha}c^{\alpha}\omega_{2}^{\alpha}g_{1}^{\alpha}(x_{1})G_{2}^{\alpha}(x_{2})}{\rho_{1}(x_{1})}.
\]
 It is clear by inspection that in fact $T_{2}(x_{1},a_{2})=a_{2}$
and $T_{2}(x_{1},b_{2})=b_{2}$, so the boundary condition is satisfied.

Similarly 
\[
T_{3}(x_{1},x_{2},x_{3})=a_{3}+L_{3}\frac{\sum_{\alpha}c^{\alpha}\omega_{3}^{\alpha}g_{1}^{\alpha}(x_{1})g_{2}^{\alpha}(x_{2})G_{3}^{\alpha}(x_{3})}{\rho_{2}(x_{1},x_{2})},
\]
 etc.

\subsection{Chebyshev polynomial case \label{sec:chebyshev}}

Consider $\alpha$ as a multi-index $\alpha=(n_{1},\ldots,n_{d})$
and define 
\[
g_{i}^{\alpha}=P_{n_{i}}\circ u_{i},
\]
 where $P_{n}$ is the $n$-th Chebyshev polynomial, and $u_{i}$
is the increasing linear map that sends $a_{i}\mapsto-1$ and $b_{i}\mapsto1$,
i.e., 
\[
u_{i}(t)=\frac{2}{b_{i}-a_{i}}t-\frac{a_{i}+b_{i}}{b_{i}-a_{i}}=\frac{t-\frac{a_{i}+b_{i}}{2}}{\frac{b_{i}-a_{i}}{2}}.
\]
 We will show in this case how to explicitly evaluate the Knothe-Rosenblatt
transport computed in Section \ref{sec:separable}. By the results
of that section, it suffices to derive an expression for the antiderivative
$G_{i}^{\alpha}$.

Now 
\[
\int P_{n}(u)\,du=\begin{cases}
u, & n=0,\\
\frac{u^{2}}{2}, & n=1,\\
\frac{1}{2(n+1)}P_{n+1}(u)-\frac{1}{2(n-1)}P_{n-1}(u), & n\geq2,
\end{cases}
\]
 and 
\[
\int P_{n}(u(t))\,dt=\int P_{n}(u)\,\frac{dt}{du}\,du.
\]
 Moreover $\frac{du_{i}}{dt}\equiv\frac{2}{b_{i}-a_{i}}$, so we can
define an antiderivative 
\[
\tilde{G}_{i}^{\alpha}(t)=\frac{b_{i}-a_{i}}{2}\begin{cases}
u_{i}(t), & n_{i}=0,\\
\frac{u_{i}(t)^{2}}{2}, & n_{i}=1,\\
\frac{1}{2(n_{i}+1)}P_{n_{i}+1}(u_{i}(t))-\frac{1}{2(n_{i}-1)}P_{n_{i}-1}(u_{i}(t)), & n_{i}\geq2.
\end{cases}
\]
 in terms of which our desired antiderivative is 
\[
G_{i}^{\alpha}(t)=\tilde{G}_{i}^{\alpha}(t)-\tilde{G}_{i}^{\alpha}(a_{i}).
\]

\subsection{Fourier mode case \label{sec:fourier}}

Consider $\alpha$ as a multi-index $\alpha=(n_{1},\ldots,n_{d})$
and define 
\[
g_{i}^{\alpha}=e_{n_{i}}\circ u_{i},
\]
 where $e_{n}(x)=e^{-2\pi inx}$ is the $n$-th Fourier mode polynomial,
and $u_{i}$ is the increasing linear map that sends $a_{i}\mapsto0$
and $b_{i}\mapsto1$, i.e., 
\[
u_{i}(t)=\frac{t-a_{i}}{b_{i}-a_{i}}.
\]
 We will repeat the computation of Section \ref{sec:chebyshev} in
this case.

Now 
\[
\int e_{n}(u)\,du=\begin{cases}
u, & n=0,\\
\frac{i}{2\pi n}e_{n}(u) & n\neq0.
\end{cases}
\]
 and 
\[
\int P_{n}(u(t))\,dt=\int P_{n}(u)\,\frac{dt}{du}\,du.
\]
 Moreover $\frac{du_{i}}{dt}\equiv\frac{1}{b_{i}-a_{i}}$, so we can
define an antiderivative 
\[
\tilde{G}_{i}^{\alpha}(t)=(b_{i}-a_{i})\begin{cases}
u_{i}(t), & n_{i}=0,\\
\frac{i}{2\pi n_{i}}e_{n}(u_{i}(t)) & n_{i}\neq0.
\end{cases}
\]
 in terms of which our desired antiderivative is 
\[
G_{i}^{\alpha}(t)=\tilde{G}_{i}^{\alpha}(t)-\tilde{G}_{i}^{\alpha}(a_{i})=(b_{i}-a_{i})\begin{cases}
u_{i}(t), & n_{i}=0,\\
\frac{i}{2\pi n_{i}}\left[e_{n}(u_{i}(t))-1\right] & n_{i}\neq0.
\end{cases}
\]

\section{Pseudospectral computations}
\label{sec:pseudocalc}

In this section we prove the claim made in Section~\ref{sec:matrixelements} that the defining property \eqref{eq:definingprop} for the matrix elements of an operator $\hat{O}$ holds for $O_{ij}$ as in \eqref{eq:altmatrix}.

First observe that for arbitrary suitably smooth $g$, we have via \eqref{eq:etaInterp} that $\langle \eta_i , g \rangle = w \, {g(\bx_{j})} / {\sqrt{J_{i}}}$. Then by replacing $g$ with $g / \sqrt{J}$ in this identity and shifting a factor of $1/ \sqrt{J}$ within the inner product, it follows that 
\begin{equation}
\label{eq:psioverlap}
\langle \psi_i, g \rangle = w \, {g(\bx_{i})} / {{J_{i}}}
\end{equation}
for all suitably smooth $g$. Hence for such $g$, we can relate the two relevant inner products via
\begin{equation}
    \langle \eta_i, g \rangle = \sqrt{J_i} \, \langle \psi_i , g\rangle,
    \label{eq:pssub}
\end{equation}
as if we were effectively identifying $\eta_i (\bx) = \psi_i (\bx) \,  \sqrt{J(\bx)} $ with $\psi_i (\bx)\, \sqrt{J_i}$.

Then for an expansion $f(\bx) = \sum_j f_j \eta_j (\bx)$ of suitably smooth $f$, we can alternatively write, following \eqref{eq:psiinterp}: $f (\bx) = \sum_j f_j \sqrt{J_j} \, \psi_j (\bx)$. Therefore $\hat{O} f (\bx) = \sum_j f_j  \sqrt{J_j}  \,  \hat{O} \psi_j (\bx)$. Then under the assumption that $\hat{O} f$ is suitably smooth, by applying \eqref{eq:pssub} with $g = \hat{O} f$ we obtain $\langle \eta_i, \hat{O} f \rangle = \sqrt{J_i} \,  \langle \psi_i, \hat{O} f \rangle  = \sum_j f_j  \sqrt{J_i J_j}  \,  \langle \psi_i, \hat{O} \psi_j \rangle$, i.e., the defining property \eqref{eq:definingprop} holds for $O_{ij}$ as in \eqref{eq:altmatrix}.

\section{Pseudopotential computations} \label{sec:PP}

In this section we discuss how to evaluate $V_I (\bx)$ using the all-electron pseudopotential of Gygi\cite{Gygi2023} in a periodic context.

The form of the potential is
\begin{align}
V_{\mathrm{PP}}(r)= & -\frac{1}{2}-\frac{\mathrm{erf}(ar)-2\left(a^{2}b+a/\sqrt{\pi}\right)re^{-a^{2}r^{2}}}{r}+\frac{\left[\mathrm{erf}(ar)-2\left(a^{2}b+a/\sqrt{\pi}\right)re^{-a^{2}r^{2}}\right]^{2}}{2} \nonumber\\
 & \quad \quad + \ \frac{\left[-2a^{2}b-4a/\sqrt{\pi}+\left(4a^{4}b+4a^{3}/\sqrt{\pi}\right)r^{2}\right]e^{-a^{2}r^{2}}}{2},
 \label{eq:pp0}
\end{align}
where $r$ is the distance from the nucleus and $a,b$ are parameters
which can be varied continuously to increase the accuracy of the potential. The PP~\eqref{eq:pp0} is directly applicable for the hydrogen atom. To obtain the PP for more general nuclear charge $Z$, one defines the
expression
\begin{equation*}
    V_{ \mathrm{PP},Z}(r)=Z^{2}V_{\mathrm{PP}}(Zr).
\end{equation*}
In this work we primarily consider $a=4$
and $b=-0.10200558466$, though we consider the impact of varying the parameter $a$ in some of our experiments. These choices are expected to yield energy error less than a few tens of $\mu E_{h}$ for molecules containing elements in the first two rows of periodic table. In terms of radially dependent functional form $V_{\mathrm{PP},Z} (r)$, we define the PP as 
\begin{equation}
    V_{I} (\bx) = V_{\mathrm{PP},Z_I} ( \vert \bx - R_I \vert ).
\end{equation}

Since we perform periodic calculations in this work, we actually need to consider the periodized pseudopotential. However simple summation in the real space diverges and in practice one does a summation in the reciprocal space and the divergent $G=0$ term is dropped. In other words the periodized potential is  
\begin{equation}
    U_I (\bx) = \sum_{\bG \neq 0} \hat{V}_I (\bG) e^{-i \bG \cdot \bx},\label{eq:FFTsum}
\end{equation}
where $\bG$ have the form $ n_1\bG_1 + n_2\bG_2  + n_3\bG_3$ of integer multiples of reciprocal lattice basis vectors, satisfying the condition that $\bG_i \cdot \bL_j = 2\pi \delta_{ij}$ where $\bL_1$, $\bL_2$, $\bL_3$ are the lattice basis vectors in real space. 

In principle one can evaluate the Fourier transform of the PP and evaluate the sum in \eqref{eq:FFTsum} using NUFFTs to calculate the periodized potential $U(\bx)$ at the grid points on the deformed coordinates. However, this is expensive because despite the fact that the PP is band-limited the number of $\bG$ vectors needed will be extremely large. To avoid this we use the Ewald summation trick whereby we split the summation into short real and reciprocal space summations. This can be done as follows
\begin{align*}
U_{I}(\bx) &= \sum_{\bG \neq 0} \hat{V}_{I}(\bG) e^{-i \bG \cdot \bx} \\
&=\sum_{\bG\neq 0}\left(\hat{V}_{I}(\bG)-\hat{V}_{1}(\bG)\right)e^{-i \bG \cdot \bx}+\sum_{\bG\neq 0} \hat{V}_{1}(\bG) e^{-i \bG \cdot \bx}\\
 & =\sum_{\bG}\left(\hat{V}_{I}(\bG)-\hat{V}_{1}(\bG)\right)e^{-i \bG \cdot \bx}+ \left(\hat{V}_{I}(0)-\hat{V}_{1}(0)\right)+\sum_{\bG\neq 0} \hat{V}_{1}(\bG) e^{-i \bG \cdot \bx}\\
 & =\sum_{\bL}\left(V_{I}(\bx+\bL)-V_{1}(\bx+\bL)\right)+\sum_{\bG\neq0}\hat{V}_{1}(\bG)e^{-i \bG \cdot \bx}-\left(\hat{V}_{I}(0)-\hat{V}_{1}(0)\right)
\end{align*}
where ${V}_{1} (\bx)$ is defined by replacing $V_{\mathrm{PP},Z} (r)$ with the softened Coulomb potential $Z\frac{\mathrm{erf}(\omega r)}{r}$. The parameter $\omega$ is chosen small enough such that
the softened potential is fairly smooth and the summation in the second term on the last line converges quickly. Meanwhile, $\hat{V}_I$ and $\hat{V}_{1}$
are respectively the Fourier transform of the PP and softened
Coulomb potential. In going from the third to the fourth line we have made use of the Poisson summation formula.

Both $V_{1}(\br)$ and $V_{I}(\br)$ converge to the usual Coulomb potential with increasing $\br$ and thus their difference goes to zero rapidly with increasing $\br$, which implies that the first summation in the last line of the equation also rapidly converges.

In the
last term both $\hat{V}_{I}(0)=4\pi\int_{0}^{\infty}V_{\mathrm{PP},Z}(r)r^{2}dr$
and $\hat{V}_{1}(0)=4\pi\int_{0}^{\infty}V_{1}(r)r^{2}dr$ individually
diverge, but their difference is bounded. We evaluate it by taking
the difference $4\pi(\int_{0}^{R}V_{\mathrm{PP},Z}(r)r^2dr-\int_{0}^{R}V_{1}(r)r^2dr)$
for a sufficiently large $R$ such that beyond this $R$ the two potentials
are equal and cancel each other. The two terms are
\begin{align*}
\int_{0}^{R}V_{\mathrm{PP}}(r)r^{2}dr= & \frac{1}{96\pi a^{3}}\left[\sqrt{\pi}\sqrt{2}\left(9\pi a^{2}b^{2}+78\sqrt{\pi}ab+49\right)\text{erf}\left(\sqrt{2}aR\right)+16\pi a^{3}R^{3}\left(\text{erf}(aR)^{2}-1\right)\right.\\
 & -48\pi a^{3}R^{2}\text{erf}(aR)-4aRe^{-2a^{2}R^{2}}\left(9\pi a^{2}b^{2}+30\sqrt{\pi}ab+17\right) \\ 
 & -48a^{3}R^{3}e^{-2a^{2}R^{2}}\left(\sqrt{\pi}ab+1\right)^{2}\\
 & \left.-32\sqrt{\pi}e^{-a^{2}R^{2}}\left(\left(3\sqrt{\pi}ab+2\right)\left(a^{2}R^{2}+1\right)\text{erf}(aR)+3a^{4}R^{3}\left(\sqrt{\pi}ab+1\right)\right)\right]\\
\int_{0}^{R}V_{1}(r)r^{2}dr= & -\frac{Z}{4}\left[2e^{-R^{2}\omega^{2}}\frac{R}{\omega\sqrt{\pi}}+\left(2R^{2}-\frac{1}{\omega^{2}}\right)\mathrm{erf}(\omega R)\right].
\end{align*}
The first term is only directly applicable for hydrogen atom, and the general case can be recoverd via $\int_{0}^{R}V_{\mathrm{PP},Z}(r)r^{2}dr=\frac{1}{Z}\int_{0}^{RZ}V_{\mathrm{PP}}(r)r^{2}dr$.

These expressions can be used to evaluate the PP $V_I (\bx_i)$ on our deformed grid
for any given parameters $a$ and $b$ defining the PP. We comment that these evaluations only have to be performed once at 
the beginning of the calculation. 

\section{Bare nuclear computations} \label{sec:bare_nuc_comp}

Our goal in this section is to compute $V_{I,i}$ as defined in \eqref{eq:VIi}. Expand this definition to obtain 
\begin{equation}
\label{eq:VIiexpand}
    V_{I,i} = -\frac{4\pi Z_I J_{i}}{w}\int\frac{1}{4\pi\vert\bx-R_{I}\vert}\,\psi_{i}(\bx)\,d\bx= 4\pi Z_I \, u_{i}(\bx_{I}),
\end{equation}
where $u_i$ is defined to be the solution of the Poisson equation
\begin{equation*}
    -\Delta u_i (\bx) = f_i (\bx) := -\frac{ J_{i}}{w} \, \psi_i (\bx).
\end{equation*}
Recall from \eqref{eq:poissonpointwise} that the solution $u_i$ satisfies
\begin{equation*}
     {J_k} \sum_j \Lap_{kj}\, u_i(\bx_j) = f_i(\bx_k) = -\frac{J_i}{w^2} \delta_{ik} ,
\end{equation*}
where we have used the property that $\psi_i (\bx_k) = w^{-1} \delta_{ik}$ in the last equality. It follows that 
\begin{equation*}
     u_i (\bx_j) =  -\frac{1}{w^2} \, \Lap^{-1}_{ij}.
\end{equation*}

Then the solution $u_i (\bx)$, evaluated at $\bx_I$, can be expanded in the $\{ \psi_j \}$ basis: 
\begin{equation}
\label{eq:uixI}
    u_i (\bx_I) = w \sum_j u_i (\bx_j)\, \psi_j (\bx_I) = -\frac{1}{w} \sum_j \Lap^{-1}_{ij} \psi_j (\bx_I).
\end{equation}
Then it follows by plugging our expression~\eqref{eq:uixI} for $u_i (\bx_I)$ into \eqref{eq:VIiexpand} and summing over $I$ that 
\begin{equation*}
    \sum_{I} V_{I,i} =  \sum_j \Lap_{ij}^{-1} b_j,
\end{equation*}
where 
\begin{equation*}
    b_i = -\frac{4 \pi }{w} \sum_I Z_I \, \psi_j (\bx_I).
\end{equation*}

In conclusion $\sum_I V_{I,i}$ can be computed for all $i$ with a single linear solve of the form $\Lap x = b$.

\end{document}